\documentclass{article}

\usepackage{PRIMEarxiv}

\usepackage[utf8]{inputenc} %
\usepackage[T1]{fontenc}    %
\usepackage{hyperref}       %
\usepackage{url}            %
\usepackage{booktabs}       %
\usepackage{amsfonts}       %
\usepackage{nicefrac}       %
\usepackage{microtype}      %
\usepackage{lipsum}
\usepackage{fancyhdr}       %
\usepackage{graphicx}       %
\graphicspath{{media/}}     %

\usepackage{natbib}
\usepackage[utf8]{inputenc} %
\usepackage[linesnumbered]{algorithm2e}
\usepackage{array}
\usepackage{algpseudocode}
\usepackage{subcaption}
\usepackage{textcomp}
\usepackage{verbatim}
\usepackage{cite}
\usepackage[T1]{fontenc}    %
\usepackage{url}            %
\usepackage{booktabs}       %
\usepackage{amsfonts}       %
\usepackage{nicefrac}       %
\usepackage{microtype}      %
\usepackage{xcolor}         %

\usepackage{amsmath}
\usepackage{amsthm}
\usepackage{bm}
\usepackage{color}
\usepackage{graphicx}
\usepackage{makecell}
\usepackage{multirow}
\usepackage{tabularx}
\usepackage{mathtools}
\usepackage{pythonhighlight}
\usepackage{listings}
\usepackage{soul}

\definecolor{pink}{rgb}{0.858, 0.188, 0.478}
\usepackage{xspace}

\newcommand{\rev}[1]{#1}

\definecolor{commentcolor}{RGB}{110,154,155}   %

\usepackage{threeparttable}
\usepackage{diagbox}

\title{GraphFusionSBR: Denoising Multi-Channel Graphs for Session-Based Recommendation
}

\author{
  Jia-Xin He, Hung-Hsuan Chen \\
  Department of Computer Science and Information Engineering \\
  National Central University \\
  Taoyuan\\
  \texttt{hohehohe1234@gmail.com, hhchen1105@acm.org} \\
}

\usepackage[title]{appendix}

\begin{document}
\maketitle

\begin{abstract}

Session-based recommendation systems must capture implicit user intents from sessions. However, existing models suffer from issues such as item interaction dominance and noisy sessions. We propose a multi-channel recommendation model, including a knowledge graph channel, a session hypergraph channel, and a session line graph channel, to capture information from multiple sources. Our model adaptively removes redundant edges in the knowledge graph channel to reduce noise. Knowledge graph representations cooperate with hypergraph representations for prediction to alleviate item dominance. We also generate in-session attention for denoising. Finally, we maximize mutual information between the hypergraph and line graph channels as an auxiliary task. Experiments demonstrate that our method enhances the accuracy of various recommendations, including e-commerce and multimedia recommendations. We release the code on GitHub for reproducibility.\footnote{\url{https://github.com/hohehohe0509/DSR-HK}}

\end{abstract}


\section{Introduction} \label{sec:intro}

Recommendation systems have become an integral part of our digital lives, influencing the way we consume information, entertainment, and products. These systems play a crucial role in various domains, including e-commerce, social media, and content streaming platforms, by tailoring user experiences based on their preferences. Traditional recommendation systems primarily rely on user profiles, past interactions, or historical browsing data to suggest items. Although effective in specific contexts, these systems face significant challenges, especially in dynamic environments where user preferences are volatile or complex to capture.

One of the main challenges in traditional recommendation systems is privacy. With increasing concerns over data protection, long-term user profiles, which are typically required by these systems, can no longer be freely utilized due to stringent privacy regulations, such as GDPR. This limits the potential of personalized recommendations and imposes restrictions on data collection practices. Additionally, the cold start problem, where a system struggles to recommend items to new users or for new items due to a lack of sufficient historical data, remains a substantial hurdle. Furthermore, user preferences can change over time, with different needs and interests emerging across different browsing sessions. Traditional models, which are based on static user profiles, often fail to capture these session-based short-term preferences.

In these contexts, session-based recommendation (SBR)~\citep{hidasi15session} has emerged as a practical approach. Unlike traditional systems, SBR models capture users' immediate preferences during a single browsing session, without requiring any long-term user information. This makes SBR inherently more privacy-friendly, as it circumvents the need for extensive user data collection. Furthermore, SBR is highly suitable for real-time recommendation tasks, where quick adaptability to evolving user interests is crucial. By focusing on short-term preferences, SBR not only mitigates the cold start problem but also likely enhances the user experience by providing timely and relevant suggestions.

Existing SBR systems can be broadly categorized into three main types: traditional models, sequence-based models, and graph neural network (GNN) models~\citep{2}. Conventional models, such as those based on Markov chains, introduce temporal dependencies by capturing transitions between adjacent items in a session. However, these models often fail to account for more complex, higher-order relationships between items. Sequence-based models~\citep{hidasi15session,li17neural,5} use recurrent neural networks (RNNs) and their variants, such as the Gated Recurrent Unit (GRU)~\citep{6}, to model sequences of items in temporal order. These models have proven effective in learning sequential patterns but may struggle with long-range or complex, high-order dependencies due to the inherent limitations of RNNs. More recently, graph neural networks (GNNs) have been increasingly applied to SBR because they allow sessions to be modeled as graphs, where nodes represent items and edges represent interactions between them. GNN-based models offer a robust framework for capturing complex item relationships, as they can naturally represent both direct and indirect dependencies. These methods can be further divided into two categories: ordinary graph-based models and hypergraph-based models. Ordinary graphs capture pairwise item interactions within a session, while hypergraphs connect all items in a session, enabling the capture of higher-order item relationships that are often crucial in modeling user intent more comprehensively.

Despite these advances, several challenges remain in the development of effective SBR systems. First, existing models often fail to leverage external information related to the items, such as knowledge graphs, which could provide valuable context by modeling semantic relationships between items. Knowledge graphs can enrich the recommendation process by incorporating domain knowledge, but they are often underutilized due to the dominance of direct item interactions. Furthermore, even when knowledge graphs are integrated, their impact can be overshadowed by noisy or irrelevant data, resulting in poor encoding of valuable external information into item representations. Lastly, noise within the session data itself -- arising from factors such as user misclicks or browsing anomalies -- can further degrade the performance of the recommendation system by introducing distractions from the user's true intent.

To address these issues, we propose a novel SBR model that leverages the strengths of multiple graph-based representations while mitigating noise. Our model, called GraphFusionSBR, introduces a multi-channel approach that integrates three types of graphs: a knowledge graph, a session hypergraph, and a session line graph. Each channel is designed to capture different aspects of user interaction. The knowledge graph channel incorporates external knowledge to enrich item representations, while a view generator dynamically removes irrelevant or redundant edges, reducing the risk of noise. In the session hypergraph channel, all session items are interconnected, enabling the model to capture complex inter-dependencies among items. To address noise within sessions, the session line graph channel assigns attention scores based on item similarity, allowing the model to focus on relevant items and generate denoised session representations. These representations are further enhanced through cross-channel contrastive learning~\citep{7}, which maximizes mutual information between the different channels, leading to more robust and accurate recommendations. 
\rev{Unlike existing multi-view approaches that simply concatenate representations from different graphs, GraphFusionSBR introduces a selective fusion mechanism centered on denoising. We argue that more information does not always lead to better recommendations; rather, filtering noise from external knowledge and internal session interactions is key.}

\rev{The main contributions of this work are summarized as follows:}

\begin{itemize}
    \item \rev{We propose a novel unified multi-channel denoising framework that integrates knowledge graphs (from knowledge base), hypergraphs (from session information), and line graphs (from session information). Crucially, we incorporate a view-generator mechanism in the knowledge graph channel and an importance extraction module in the line graph channel to actively filter out irrelevant semantic edges and noisy session interactions before fusion.}
    
    \item \rev{We bridge the gap between semantic-based reasoning (via knowledge graph) and structure-based reasoning (via hypergraph and line graph). By maximizing mutual information between these views through cross-channel contrastive learning, we enhance the robustness of item representations against data sparsity.}

    \item \rev{Extensive experiments on three benchmark datasets (Tmall, RetailRocket, and KKBox) demonstrate that GraphFusionSBR consistently outperforms state-of-the-art baselines. Furthermore, our theoretical analysis confirms that this performance gain comes with linear computational complexity, making it scalable for real-world applications.}
\end{itemize}

The remainder of the paper is organized as follows. In Section~\ref{sec:rel-work}, we review previous work on session-based recommendation, knowledge graph, hypergraph, and hypergraph-based recommendation, and self-supervised learning on recommendation systems. Section~\ref{sec:method} introduces our proposed method. Section~\ref{sec:exp} shows the results of the comparison of GraphFusionSBR with baseline methods based on three datasets. Finally, we conclude and discuss future work in Section~\ref{sec:disc}.

\section{Related Work} \label{sec:rel-work}

\rev{This section reviews the literature relevant to our proposed method. We first discuss the evolution of session-based recommendation systems in Section~{\ref{sec:session-based-review}}. Next, we examine the utilization of knowledge graphs (KGs) in recommendations in Section~{\ref{sec:kg-review}}. In Sections~{\ref{sec:hypergraph-review}} and {\ref{sec:ssl-review}}, we review hypergraph approaches and self-supervised learning. Finally, we discuss existing multi-view fusion strategies and distinguish our approach in Section~{\ref{sec:cmp-review}}.}

\subsection{Session-Based Recommendation Systems} \label{sec:session-based-review}

Session-based recommendation systems can be categorized into three types. First, traditional recommendation systems mainly use Markov chains to extract temporal information. FPMC~\citep{11} combines Markov chains with matrix factorization techniques for a sequential recommendation, considering both short-term dynamic sequences and long-term preference features. However, Markov chains only consider transitions between adjacent states, limiting their representation capabilities for tasks requiring high-order structure modeling. Second, sequence-based recommendation systems frequently employ recurrent neural networks (RNNs) to capture the sequential relationships between items. SBR-RNN~\citep{hidasi15session} is an early study to apply RNNs to SBR, introducing ranking loss and session-parallel training for more accurate predictions. Subsequent studies, such as NARM~\citep{li17neural}, used hybrid encoders and attention mechanisms to model users' sequential behaviors and capture the primary purpose of users in the current session. Third, GNN-based recommendation systems, which learn item transitions through session graph structures. SR-GNN~\citep{wu19session} was an early work that converted session sequences into session graphs, utilizing gating mechanisms to control and update the message flow. GCE-GNN~\citep{wang20global} combines graph convolution and self-attention, considering both local node structures and global graph information to generate richer representations for each node. Although GNN methods usually outperform RNNs, they primarily use ordinary graphs constructed by pairing items to represent relationships, which may limit their ability to capture higher-order item relationships.

\subsection{Knowledge Graphs} \label{sec:kg-review}

To address the issue of data sparsity in recommendation systems, knowledge graphs have been proposed as a solution. A knowledge graph is a structured knowledge base that represents semantic relationships between entities. Approaches to integrate knowledge graphs into recommendation systems include embedding-based methods (e.g., TransE~\citep{17} and TransR~\citep{18}), path-based methods (e.g., meta-path~\citep{29} and meta-graph~\citep{30}), and GNN-based methods (e.g., KGAT~\citep{19} and KGCN~\citep{20}).

\rev{However, directly integrating KGs into SBR poses challenges. GNN-based methods, such as KGAT~{\citep{19}} and KGCN~{\citep{20}}, have shown promise in propagating information through neighbors. Yet, these methods often suffer from ``knowledge noise,'' where irrelevant connections in the KG dilute the user's core intent. Unlike previous works that treat all KG edges equally important, our approach incorporates a denoising mechanism to selectively sample high-value edges, ensuring that only relevant external knowledge enhances the session representations.}

\subsection{Hypergraph and Recommendation} \label{sec:hypergraph-review}

Hypergraphs extend traditional graphs by allowing edges to connect more than two nodes, thus capturing higher-order relationships. With the success of GNNs in graph data tasks, researchers have started to focus on extending these techniques to hypergraphs. HGNN~\citep{8} and HyperGCN~\citep{9} are pioneering works that apply graph convolutional networks to hypergraphs, proposing methods for learning hypergraph representations through hyperedge-average-based hypergraph convolution and hypergraph Laplacian.

Some studies combine hypergraphs with recommendation systems. For example, HIDE~\citep{10} designed three types of hyperedges to convert each session into a hypergraph. However, constructing separate hypergraphs for different sessions may fail to capture the transitions and connections of user interests across sessions. Additionally, the lack of temporal information modeling may limit the model's capacity to capture the evolution of dynamic user interest.

\subsection{Self-Supervised Learning in Recommendation Systems} \label{sec:ssl-review}

Self-supervised learning (SSL) learns data representations from raw, unlabeled data. Initially used in computer vision and natural language processing, SSL utilizes data augmentation techniques such as rotation and cropping. Recently, SSL has been applied to GNNs, for example, by maximizing the mutual information between local node features and global graph representations to improve node representation learning.

Some work attempts to integrate SSL with sequential recommendation tasks, using item correlations and other self-supervised strategies to extract effective self-supervised signals from raw data, enhancing model performance. However, existing SSL methods primarily focus on improving representation learning capabilities, while overlooking significant challenges in recommendation systems, such as cold start, data sparsity, and modeling high-order relationships. In SBR, the inherent data sparsity makes random dropout-based self-supervised signal generation strategies less effective.

Researchers have combined SSL with contrastive learning (CL) and multi-view semi-supervised learning to address data sparsity in new scenarios. Representative works include KGCL~\citep{kgcl2022}, which augments data based on knowledge graphs to improve the quality of representation. \rev{However, while these SSL-based methods improve robustness against sparse data, they often focus on a single aspect of data augmentation and may struggle when facing significant noise or when integrating heterogeneous information sources.}

\subsection{Multi-view and Graph Fusion Approaches} \label{sec:cmp-review}

\rev{Recent trends in SBR have started studying fusing multiple views (e.g., sequence view and graph view) to enrich representations. Prominent methods, such as COTREC~{\citep{23}}, utilize co-training across different graph views (e.g., session graphs and item graphs) to enforce consistency and maximize mutual information. However, most existing fusion frameworks rely on simple concatenation or soft attention mechanisms that may not effectively handle the noise introduced by heterogeneous channels. Unlike these approaches, our GraphFusionSBR differs by explicitly integrating a ``denoising-first'' strategy within each channel (knowledge graph and line graph) before the fusion stage, ensuring that the collaborative learning process is driven by high-quality, refined signals rather than raw, noisy data.}

\section{Method} \label{sec:method}

\rev{In this section, we detail the proposed GraphFusionSBR framework. We first present the problem definition and graph construction in Section~{\ref{sec:problem-def-graph-construct}}. Then, we describe the three core channels -- knowledge graph channel, hypergraph channel, and line graph channel -- along with their denoising and encoding mechanisms in Section~{\ref{sec:model-archi}}. Finally, we provide a theoretical complexity analysis of our model in Section~{\ref{sec:complexity}}.}

\subsection{Problem Definition and Graph Construction} \label{sec:problem-def-graph-construct}

\begin{table}[tbp]
\centering
\caption{Key Symbol Glossary}
\label{tab:symbols}
\begin{tabular}{ll}
\toprule
\textbf{Symbol} & \textbf{Description} \\
\midrule
$ I $ & Set of items in the recommendation system \\
$ N $ & Total number of items \\
$ s $ & A session, i.e., an ordered sequence of item interactions \\
$ m $ & Number of interactions (clicks) in a session \\
$ x_i \in \mathbb{R}^d $ & Embedding vector of item $ i $ \\
$ X \in \mathbb{R}^{N \times d} $ & Matrix of all item embeddings \\
\midrule
$ \mathcal{G}_h = (V, E) $ & Hypergraph with nodes $ V $ (items) and hyperedges $ E $ \\
$ H $ & Incidence matrix of the hypergraph \\
$ D $ & Diagonal matrix of node degrees in the hypergraph \\
$ B $ & Diagonal matrix of hyperedge degrees in the hypergraph \\
\midrule
$ \mathcal{G}_l = (V_l, E_l) $ & Line graph with nodes $V_l$ (sessions) and edges $E_l$ \\
$ W_{p,q} $ & Ratio of shared nodes between hyperedges $e_p$ and $e_q$ in $\mathcal{G}_h$ \\
\midrule
$ \mathcal{G}_k = \{(h, r, t)\} $ & Knowledge graph represented as triples (head, relation, tail) \\
$ w_e $ & Importance weight for edge $ e $ in the knowledge graph channel \\
$ p_e $ & Sampling probability for edge $ e $, computed via Gumbel-Max \\
$ \lambda $ & Uniform random variable in $ [0,1] $ (used in reparameterization) \\
$ \tau_b $ & Temperature parameter for the Gumbel distribution \\
\midrule
$ \theta_k $ & Session embedding from the knowledge graph channel \\
$ \theta_h $ & Session embedding from the hypergraph channel \\
$ \theta_l $ & Session embedding from the line graph channel \\
\midrule
$ L_{rec} $ & Recommendation loss (cross-entropy) \\
$ L_{ssl} $ & Self-supervised (contrastive) loss \\
$ L_{KG} $ & Knowledge graph loss (TransR ranking loss) \\
$ L $ & Overall loss: $L = L_{rec} + \lambda_1 L_{ssl} + \lambda_2 L_{KG}$ \\
\bottomrule
\end{tabular}
\end{table}

This section introduces the foundational concepts and objectives of the proposed GraphFusionSBR model, with a focus on the multiple graph structures used to enhance session-based recommendations. Table~\ref{tab:symbols} shows the key symbols used in the paper.

We define the item set in the recommendation system as $I = \{i_1, i_2, \ldots, i_{N-1}, i_N\}$, where $N$ denotes the total number of items available. Each session, represented as $s = \{i_{s,1}, i_{s,2}, \ldots, i_{s,m-1}, i_{s,m}\}$, includes a sequence of interactions with items, where $m$ is the number of clicks, and each item $i_{s,k} \in I$ ($1 \leq k \leq m$) reflects an interaction with the session $s$. To support learning from these sessions, each item $i \in I$ is embedded into a vector space, with its representation denoted as $x_i \in \mathbb{R}^d$. The collective embeddings of all items in the system are represented as $X \in \mathbb{R}^{N \times d}$. The primary objective of GraphFusionSBR is to accurately predict the next likely item $i_{s,m+1}$ in a session $s$, using multi-channel information from hypergraphs, line graphs, and knowledge graphs.

\subsubsection{Hypergraph}
We represent the hypergraph by $\mathcal{G}_h = (V, E)$, where $V$ is a set of $N$ distinct nodes (corresponding to items) and $E$ is a set of $M$ hyperedges. Each hyperedge $\epsilon \in E$ links multiple nodes, providing a means to capture higher-order relationships between items that co-occur within a session. In this model, each hyperedge $\epsilon$ is assigned a positive weight $W_{\epsilon\epsilon}$, set to 1 in this study. The matrix $\boldsymbol{W}$ is an $M \times M$ diagonal matrix containing the hyperedge weights. The hypergraph structure is represented by an incidence matrix $\boldsymbol{H}$ of size $N \times M$, where $H_{i\epsilon} = 1$ if the hyperedge $\epsilon$ contains node $v_i$, and $H_{i\epsilon} = 0$ otherwise. 
\rev{The degrees of nodes and hyperedges are calculated based on standard definitions to normalize the graph structure, as detailed in Equation~{\ref{eq:hg-degree}} in Appendix~\mbox{\ref{app:hyper-conv}}. This structure allows the system to learn from item co-occurrences.}

\subsubsection{Line Graph}\label{Line graph}

Given a hypergraph $\mathcal{G}_h = (V, E)$, we construct its line graph $\mathcal{G}_l = (V_l, E_l)$, which redefines hyperedges as nodes. Here, each node $v_e \in V_l$ represents a hyperedge $e \in E$ in $\mathcal{G}_h$. If two hyperedges $e_p$ and $e_q$ share at least one node in $\mathcal{G}_h$, an edge is created between $v_{e_p}$ and $v_{e_q}$ in $\mathcal{G}_l$. Formally:
\begin{align*}
    V_l &= \{v_e : e \in E\}, \\
    E_l &= \{(v_{e_p}, v_{e_q}) : e_p, e_q \in E, |e_p \cap e_q| \geq 1\}.
\end{align*}
Each edge $(v_{e_p}, v_{e_q}) \in E_l$ is assigned a weight $W_{p,q} = |e_p \cap e_q| / |e_p \cup e_q|$, indicating the ratio of shared nodes between the hyperedges $e_p$ and $e_q$ in $\mathcal{G}_h$. The line graph effectively captures co-occurrence structures between sessions, improving the model's understanding of session interrelationships through shared items.

\subsubsection{Knowledge Graph}

A knowledge graph $\mathcal{G}_k = \{(h, r, t)\}$ is represented as a collection of triples $(h, r, t)$, where $r \in \mathcal{R}$ defines the relationship between a head entity $h \in \mathcal{E}$ and a tail entity $t \in \mathcal{E}$. Here, $\mathcal{E}$ and $\mathcal{R}$ represent the sets of entities and relationships, respectively. 

The entity set $\mathcal{E} = \mathcal{I} \cup \mathcal{A}$ combines item entities $\mathcal{I}$ (such as movies, books) with attribute entities $\mathcal{A}$ (such as authors, genres). Relations in $\mathcal{R}$ describe the connections between entities, such as ``written by author'' or ``belongs to genre.'' For example, a triple like (Harry Potter, author, J.K. Rowling) indicates that ``J.K. Rowling'' is the author of the book ``Harry Potter.'' The relation set $\mathcal{R}$ includes both forward and reverse connections (e.g., ``written by'' and ``wrote''), thereby facilitating comprehensive semantic relationships. This knowledge graph enriches item embeddings by integrating domain-specific knowledge, thereby improving the model's recommendation precision and interpretability.

GraphFusionSBR enhances the context of each session by providing additional insights into items and their relationships based on the structured information within the knowledge graph. This multi-layered representation of knowledge enables the recommendation system to generalize more effectively across different types of content, thereby improving both accuracy and relevance in predictions.

\subsection{Model Architecture} \label{sec:model-archi}

\begin{figure*}[tb]
    \centering
    \includegraphics[width=\textwidth]{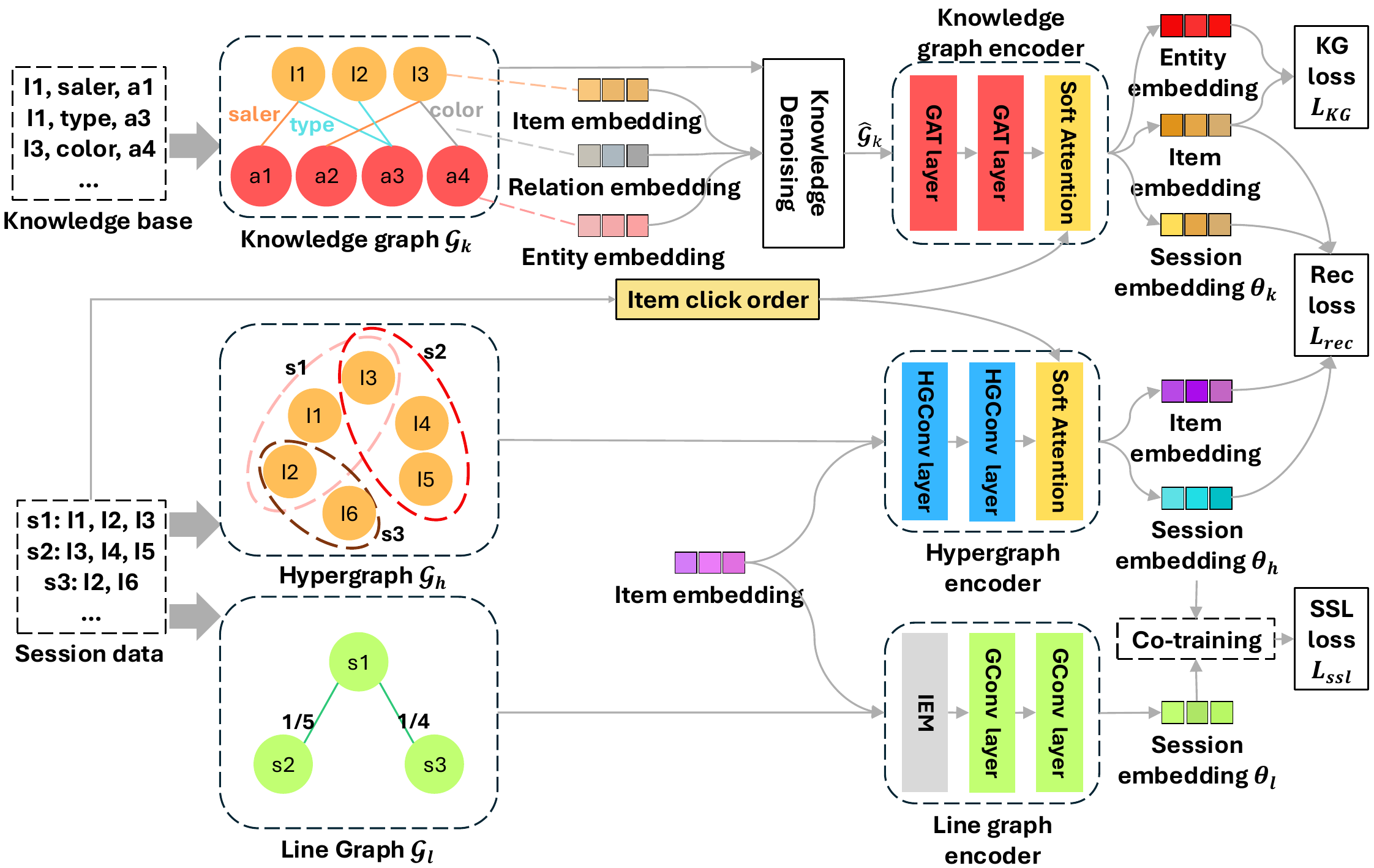}
    \caption{\rev{The overall architecture of GraphFusionSBR. The raw session data and knowledge base are processed through three parallel channels. The knowledge graph channel (top) integrates a knowledge base and applies a denoising view generator. The hypergraph channel (middle) captures high-order correlations. The line graph channel (bottom) extracts cross-session information via an Importance Extraction Module (IEM). The final representations are fused for next-item prediction, optimized jointly by recommendation loss, cross-channel contrastive loss, and knowledge graph loss.}}
    \label{fig:archi}
\end{figure*}

\rev{As illustrated in Figure~{\ref{fig:archi}}, the proposed GraphFusionSBR architecture processes raw session data and knowledge base through a multi-channel pipeline to generate final predictions. The data flow operates as follows.

First, the input stage transforms the raw session sequences and the external knowledge base into three distinct graph structures: a knowledge graph constructed from item attributes (top), a session-based hypergraph (middle), and a session-to-session line graph (bottom).

Next, these inputs are processed in parallel through three specific encoders.
}
\begin{itemize}
    \item \rev{In the knowledge graph channel, the graph structure first passes through a denoising module to filter irrelevant edges before being encoded by GAT layers to generate semantic-enriched item and session embeddings, which are regularized by a translation-based knowledge graph loss ($L_{KG}$).}

    \item \rev{In the hypergraph channel, session items are constructed into hyperedges and processed by hypergraph convolutional layers to capture high-order intra-session correlations.}

    \item \rev{In the line graph channel, an Importance Extraction Module (IEM) initializes session features, which Graph Convolutions then refine to capture inter-session dynamics.}
\end{itemize}

\rev{Finally, in the aggregation and prediction stage, the item and session representations from the knowledge graph and hypergraph channels are concatenated to compute the recommendation score (Rec Loss $L_{rec}$). Simultaneously, the session representations from the hypergraph and line graph channels serve as two views for contrastive learning (SSL Loss $L_{ssl}$), enforcing cross-view consistency. The entire model is jointly optimized by these objectives along with the knowledge graph loss ($L_{KG}$) computed within the KG channel.}

\subsubsection{Knowledge Graph Channel}

The knowledge graph channel utilizes the knowledge graph and session items to generate entity embeddings, item embeddings, and session embeddings.  This channel consists of a knowledge denoising module, a knowledge graph encoder, and a reverse position embedding module.

The knowledge denoising module denoises unrelated information from the knowledge graph by adopting a view generator~\citep{32}, which models an importance weight $w_e$ for each edge $e$ and generates a soft mask probability $p_e \in [0, 1]$. The edge $e$ is retained with probability $p_e $. The weight $w_e$ is computed by:
\begin{equation}
    w_e = MLP([x^{(0)}_{k,I} || x^{(0)}_{k,A}])
\end{equation}
where edge $e = (I, A)$ connects an item $I$ and an attribute $A$, $MLP$ is a multi-layer perceptron, and $x^{(0)}_{k,I}$ and $x^{(0)}_{k,A}$ are the embeddings of the item and attribute entities in the knowledge graph channel. Higher $w_e$ scores indicate critical edges that should be retained. We then compute the sampling probability using $w_e$, employing the Gumbel-Max reparameterization trick~\citep{25,26} to ensure $p_e \in [0, 1]$ is differentiable:
\begin{equation}
    p_e = \sigma((\log(\lambda) - \log(1 - \lambda) + w_e) / \tau_b)
\end{equation}
where $\lambda \sim \text{Uniform}(0, 1)$, $\sigma(\cdot)$ is sigmoid, and the hyperparameter $\tau_b$ controls the steepness of the Gumbel distribution. Higher $\tau_b$ values make the distribution smoother, whereas lower $\tau_b$ values make it steeper, approximating a discrete distribution. This reparameterization not only approximates sampling from a discrete probability distribution but also introduces randomness to discover new possibilities and avoid local optima. The denoised knowledge graph is denoted as $\hat{\mathcal{G}_k}$. 

\rev{After obtaining $\hat{\mathcal{G}}_{k}$, we employ a knowledge graph encoder based on a Graph Attention Network (GAT) to aggregate neighbor information. Specifically, attention coefficients and node feature updates follow the standard GAT formulation, which is provided in Appendix~\mbox{\ref{app:gat}}.}

After multiple layers of aggregation, we average the item embeddings from all layers to obtain the final item embedding: $x^k_i = \dfrac{1}{L+1} \sum^L_{l=0} x^{k(l)}_i$.

To incorporate the position information of an item into a session, we adopt reverse position embedding using a learnable position vector matrix $P_r = [p_1, p_2, \ldots, p_m]$. Eventually, for an item $t$ in session $s = [i_{s,1}, \ldots, i_{s,m}]$, its representation is:
\begin{equation}\label{position}
    x^{k*}_t = \tanh(W_1 [x^k_t || p_{m-t+1}] + b)
\end{equation}
where $W_1 \in \mathbb{R}^{d \times 2d}$, $p_{m-t+1} \in \mathbb{R}^d$, and $b \in \mathbb{R}^d$ are learnable parameters.

The session embedding is derived by integrating item representations $x^{k*}_t$ in the session. We use soft attention~\citep{wu19session} to generate session embedding $\theta_k$:
\begin{equation}
     \theta_k = \sum\limits_{t=1}^{m} \alpha^k_t x^k_t = \sum_{t=1}^m f^T \sigma(W_2 x^k_s + W_3 x^{k*}_t + W_4 x^k_{\text{last}} + c) x^k_t,
\end{equation}
where $x^k_s$ is the average item embedding in the session, representing the context information of the session, $x^{k*}_t$ is the embedding of the item $t$ in the session, $x^k_{\text{last}}$ is the last clicked item in the session, which is considered to be the most relevant item of current interest. The parameters $f^T \in \mathbb{R}^d$, $W_2 \in \mathbb{R}^{d \times 2d}$, $W_3 \in \mathbb{R}^{d \times 2d}$, $W_4 \in \mathbb{R}^{d \times 2d}$, and $c \in \mathbb{R}^d$ are attention parameters. After obtaining the session embedding $\theta_k$ using the knowledge graph, it will be used to optimize the recommendation loss in Section~\ref{sec:Lr}.

\subsubsection{Hypergraph Channel}

The hypergraph channel utilizes sessions to generate a hypergraph, where each node represents an item, and each hyperedge connects all items within a session. This channel uses a hypergraph convolutional network (HGCN) to encode nodes and hyperedges. \rev{To propagate item embeddings through the hypergraph structure, we adopt the standard hypergraph convolution operation defined in Equation~{\ref{eq:hg-conv}} in Appendix~\mbox{\ref{app:hyper-conv}}.}

Conceptually, this convolution performs a ``node-to-hyperedge-to-node'' feature transformation, allowing the model to capture complex high-order correlations.After multiple convolutional layers, we average the item embeddings to obtain the final item embeddings $X_h$. Finally, we apply reverse positional embedding and soft attention to aggregate these item representations into the session embedding $\theta_h$ for the hypergraph channel.

\subsubsection{Line Graph Channel}

The line graph channel creates a graph where each node represents a session, and nodes are connected if they share items, with edge weights based on the Jaccard similarity of the item sets. This channel includes an importance extraction module (IEM) and a line graph encoder. Although a hypergraph captures complex relationships within sessions, the sparsity of session data and noise might affect its effectiveness. Inspired by COTREC~\citep{23}, we use self-supervised signals to enhance predictions.

The IEM generates the embedding of a session $S= \{x^{(0)}_1, x^{(0)}_2, \ldots, x^{(0)}_t\}$ by a weighted sum $x^{(0)}_i$s ($x^{(0)}_i$ is the initial embedding of the item $i$). We estimate the importance of each item using the self-attention mechanism in SR-IEM~\citep{24}. First, we calculate the similarity matrix $C$:
\begin{equation}
    C = \dfrac{\sigma(QK^T)}{\sqrt{d}} = \dfrac{\sigma(\sigma(W_q S) \sigma(W_k S)^T)}{\sqrt{d}},
\end{equation}
where $Q$ and $K$ are the queries and keys used in attention, $W_q \in \mathbb{R}^{d \times l}$ and $W_k \in \mathbb{R}^{d \times l}$ are trainable parameters.

We compute the average similarity of an item $i$ to other items in the session as the item's importance score $\alpha_i$.

\begin{equation}
    \alpha_i = \dfrac{1}{t} \sum\limits_{j=1, j \neq t}^{t} C_{ij}
\end{equation}
where $C_{ij}$ is the $(i,j)$th entry in $C$. We apply a softmax layer to obtain the final importance $\beta = \text{softmax}(\alpha)$. 

Finally, the embedding for a specific session $s$, denoted as $\theta_{l,s}^{(0)}$, is computed as:
\begin{equation}
\theta_{l,s}^{(0)}=\sum_{i=1}^{t}\beta_{i}x_{i}^{(0)} \quad (11)
\end{equation}

We simply stack the embeddings of all $M$ sessions to form the initial node feature matrix of the line graph, denoted as $\Theta_{l}^{(0)} \in \mathbb{R}^{M \times d}$.
Then, we convolve the line graph $\mathcal{G}_{l}$ to aggregate information by

\begin{equation}
    \Theta_l^{(l+1)} = \hat{D}^{-1} (A+I) \Theta_l^{(l)},
\end{equation}
where $A$ is the adjacency matrix of the line graph, $I$ is the identity matrix, and $\hat{D} \in \mathbb{R}^{M \times M}$ is a diagonal degree matrix with $\hat{D}_{p,p} = \sum^M_{q=1} \hat{A}_{p,q}$.

We average the session embeddings of all layers to obtain the final session embedding $\Theta_l$, i.e.,  $\Theta_l = \dfrac{1}{L+1} \sum^L_{l=0} \Theta_l^{(l)}$, where each row represents a session's embedding $\theta_l$.

\subsubsection{SSL Loss Across Channels}

The hypergraph and line graph obtain intra-session and cross-session representations, respectively, so co-training the two channels may capture complementary information~\citep{23}. Specifically, for a session $p$ in the line graph, we use item representations from the hypergraph encoder to predict the next item in session $p$:
\begin{equation}
    y^p_h = \text{Softmax}(X_h \theta^p_h)
\end{equation}
where $\theta^p_h$ is the session embedding in the hypergraph, and $y^p_h \in \mathbb{R}^N$ represents the probability that each item is recommended to the session $p$ in the hypergraph.

Using the computed probabilities, we select the top-$K$ items as positive samples $c^{p+}_l$, which will be used later by the line graph encoder. For negative samples $c^{p-}_l$, we select the $10\%$ most related items and randomly choose $K$ items as negative samples, excluding the selected positive samples. These more challenging negative samples provide the model with more information to distinguish between positive and negative samples. Similarly, we select positive samples $c^{p+}_h$ and negative samples $c^{p-}_h$ for the hypergraph encoder. Given the positive and negative samples, we maximize the following self-supervised (SSL) loss~\citep{33}:

\begin{equation}
\label{loss:ssl}
\begin{split}
    L_{ssl} = -\log \dfrac{\sum_{i \in c^{p+}_h} \psi(x^{\text{last}}_h, \theta^p_h, x^i_h)}{\sum_{i \in c^{p+}_h \cup c^{p-}_l} \psi(x^{\text{last}}_h, \theta^p_h, x^i_h)}
    -\log \dfrac{\sum_{i \in c^{p+}_l} \psi(x^{\text{last},(0)}_{h}, \theta^p_l, x^{i,(0)}_{h})}{\sum_{i \in c^{p+}_l \cup c^{p-}_l} \psi(x^{\text{last},(0)}_{h}, \theta^p_l, x^{i,(0)}_{h})},
\end{split}
\end{equation}
where $x^{\text{last}}_h$ is the embedding of the last clicked item in the hypergraph channel, $\theta^p_h$ and $\theta^p_l$ represent the session embeddings for session $p$ from the hypergraph and line graph channels respectively, $x^i_h$ and $x^{i,(0)}_{h}$ denote the final and initial embeddings of the item $i$ in the hypergraph channel, respectively, $\psi(a_1, a_2, a_3) = \exp(f(a_1 + a_2, a_2 + a_3) / \tau)$, $\tau$ is the temperature parameter, and $f(\cdot)$ computes the cosine similarity between two vectors.

Through this contrastive learning method, the two encoders exchange information, enhancing session and item representations.

\subsubsection{Knowledge Graph Loss}
To encode the knowledge graph, we utilize TransR~\citep{18} to optimize the embeddings of the entities and relations. 
\rev{We calculate the plausibility score $g(h,r,t)$ for each triple, which measures the distance between the head and tail entities in the relation space (see Appendix~\mbox{\ref{app:transr-score}} for the detailed scoring function).}

We optimize using pairwise ranking loss:
\begin{equation}
    L_{KG}(h, r, t^+, t^-) = -\log \sigma(g(h, r, t^-) - g(h, r, t^+)),
\end{equation}
where $(h, r, t^+)$ is a true triple, and $(h, r, t^-)$ is a false triple created by randomly replacing an entity.

\subsubsection{Recommendation Loss}\label{sec:Lr}

To predict the probability $\hat{y}_i$ that an item $i$ is the next item in the session $s$, we concatenate the session/item embeddings from the hypergraph and knowledge graph channels and use the dot product to compute the score $z_i$ for item $i$:

\begin{equation}
    z_i = (\theta_h || \theta_k)^T (x_{h,i} || x_{k,i})
\end{equation}

We apply the softmax function to obtain the predicted probability $\hat{y}_i$:

\begin{equation}
\hat{y}_i = \frac{\exp(z_i)}{\sum_{j=1}^{N} \exp(z_j)}
\end{equation}

Finally, we apply the cross-entropy as the recommendation loss function:
\begin{equation}
    L_{rec} = -\sum\limits_{i=1}^{N} I(y_i) \log(\hat{y_i}) + (1-I(y_i)) \log(1-\hat{y_i}),
\end{equation}
where $I(y_i)$ is a boolean indicating whether item $i$ is the next item. We use the Adam optimizer to minimize $L_{rec}$.

\subsubsection{Overall Loss and the Training Process}
Finally, the overall loss function is:
\begin{equation}
    L = L_{rec} + \lambda_1 L_{ssl} + \lambda_2 L_{KG},
\end{equation}
where $\lambda_1$ and $\lambda_2$ balance different losses. 

During training, we first optimize the knowledge graph loss, followed by recommendation and contrastive learning losses. This approach integrates valuable knowledge into the model to support the recommendation task. Algorithm~\ref{alg:train} in Appendix~\ref{app:train-algo} shows the entire training process.

\subsection{Complexity Analysis} \label{sec:complexity}

\rev{To assess the scalability of GraphFusionSBR, we analyze the time complexity of its key components. Let $N$ be the number of items, $M$ be the number of sessions (hyperedges), and $d$ be the embedding dimension. Below, we analyze the complexity of the hypergraph channel, the knowledge graph channel, and the line graph channel.

The complexity of hypergraph convolution is dominated by the incidence matrix multiplications, which is $O(L_{HG} \times (|V| + |E_{HG}|) \times d)$, where $L_{HG}$ is the number of layers and $|E_{HG}|$ is the number of non-zero entries in the incidence matrix. This is linear with respect to the number of interactions in the training data.

For the knowledge graph channel, the GAT-based encoder operates on the KG edges. The complexity is $O(L_{KG} \times |E_{KG}| \times d)$, where $L_{KG}$ is the number of layers and $|E_{KG}|$ is the number of edges in the pruned KG. The view-generator adds a small overhead of $O(|E_{KG}| \times d)$ but significantly reduces the graph size for subsequent steps.

The construction of the line graph can be pre-computed. The cost of graph convolution is $O(L_{LG} \times |E_{LG}| \times d)$, where $L_{LG}$ is the number of layers and $|E_{LG}|$ is the number of edges in the line graph. This is proportional to the number of connections of shared elements between sessions.

Overall, the training complexity of GraphFusionSBR is linear with respect to the scale of the graph (nodes and edges). While the multi-channel architecture introduces additional parameters compared to single-view models (e.g., SR-GNN), its complexity remains comparable to that of other state-of-the-art multi-view methods, ensuring the model is computationally feasible for large-scale datasets.}

\section{Experiments} \label{sec:exp}

\begin{table*}[tbh]
\centering
\caption{Dataset statistics. Entities include both item entities and attribute entities.} \label{tab:Statistics}
\begin{tabular}{crrr}
        \toprule
         & Tmall & RetailRocket & KKBox \\
        \midrule
        Training Sessions & 348,766 & 388,541 & 432,689 \\
        Testing Sessions & 25,412 & 3,469 & 81,824 \\
        Items & 41,512 & 25,390 & 23,838 \\
        Relations & 3 & 7 & 10 \\
        Triples & 124,506 & 166,355 & 178,705 \\
        Entities & 50,840 & 73,946 & 60,826 \\
        Average Session Length & 6.69 & 6.03 & 12.33 \\
        \bottomrule
    \end{tabular}
\end{table*}

In this section, we present a comprehensive evaluation of GraphFusionSBR on three benchmark datasets: Tmall, RetailRocket, and KKBox. We detail the dataset preprocessing, hyperparameter configurations, and experimental protocols used to assess the model's performance. Furthermore, we compare GraphFusionSBR with several state-of-the-art baselines using standard metrics such as Precision@$K$ and MRR@$K$. Our experiments include ablation studies and hyperparameter analyzes to demonstrate the contribution and robustness of each component within our multi-channel framework.

\subsection{Datasets, Preprocessing, and Hyperparameters}

We use three datasets for our experiments: Tmall, RetailRocket, and KKBox. The Tmall dataset, sourced from the IJCAI-15 competition, contains anonymized user shopping logs from the Tmall online shopping platform, providing a diverse set of interactions suitable for evaluating session-based recommendations. The RetailRocket dataset, obtained from a Kaggle competition, includes six months of user browsing activity from an e-commerce company. Finally, the KKBox dataset captures users' music listening behavior over four months, including playlist information, which helps to understand more extended sequences in session data. Each dataset is preprocessed distinctly to suit the specific characteristics of the recommendation system and improve model training.

To standardize our experiments, we adopt a consistent preprocessing approach based on previous studies~\citep{wang20global,wu19session} across all datasets. First, we filter out all sessions of length one and items that appear fewer than 5 times to focus on commonly interacted items. Crucially, we apply data augmentation via sequence splitting. For each session $s=[i_{s,1}, i_{s,2}, ..., i_{s,m}]$, we generate a series of labeled sequences to train the model on intermediate states: $([i_{s,1}], i_{s,2})$, $([i_{s,1}, i_{s,2}], i_{s,3})$, ..., $([i_{s,1}, i_{s,2}, ..., i_{s,m-1}], i_{s,m})$. This technique significantly increases the volume of training data, helping the model learn and predict the immediate following items.

For the Tmall dataset, we follow the setup of previous research~\citep{wang20global}. We select the first 120,000 sessions to form our primary dataset. Applying the aforementioned sequence splitting, this results in 348,766 labeled training sequences, as shown in Table~\ref{tab:Statistics}. Sessions in the last ten months are used as the test set to evaluate the model's prediction performance in more recent sessions. To enrich the recommendation system with contextual knowledge, we incorporate Tmall's public attribute dataset, which provides additional attributes such as the seller, main item category (e.g., clothing, jewelry), and sub-item category (e.g., tops, dresses). This knowledge graph enhances the model's predictive power and facilitates personalized recommendations by linking items with shared attributes and categories, thereby improving the interpretability of the model's decisions.

In the case of the RetailRocket dataset, the sessions are mostly short, with an average length of just over two items. Therefore, to create a balanced dataset, we retain 10,000 sessions of lengths 2-4 and all sessions longer than 4, resulting in a dataset of approximately 40,000 sessions, which are further augmented through sequence splitting. The final four months of data serve as the test set, simulating a realistic scenario of predicting future user interactions. RetailRocket provides static (e.g., category, seller) and dynamic (e.g., price, inventory) attributes. We use only static attributes to construct the knowledge graph for simplicity and consistency, as dynamic attributes can introduce noise when predicting long-term preferences. To ensure consistency of attributes across items, we select only those attributes that are present in at least $90\%$ of the items.

For the KKBox dataset, only tracks that users listened to for at least a minute are retained to ensure meaningful engagement in each session. Additionally, because the removal of certain tracks could disrupt the continuity of interest within sessions (for example, a session initially containing $[i_1, i_2, ..., i_{20}]$ could become $[i_1, i_{20}]$ after filtering), we keep only those sessions where the order difference between the remaining tracks is no greater than 3, preserving the original flow of user interest. The knowledge graph is constructed similarly to that of the Tmall dataset, allowing for a structured representation of music attributes, such as genre and artist, which enhances the relevance and interpretability of the recommendations. Table~\ref{tab:Statistics} provides a detailed overview of the final statistics for the three datasets.

For hyperparameters, we align with previous research guidelines~\citep{32}, setting the embedding size to 112 to capture sufficient representational detail while striking a balance between computational efficiency. We use a batch size of 100 to ensure efficient GPU utilization and an initial learning rate of 0.001 to ensure steady convergence during training. The model architecture comprises a single layer, simplifying the structure while maintaining accuracy. We set the number of positive and negative samples, $K$, to 5 to balance precision and diversity in contrastive learning. Baseline models are configured with their default hyperparameters as reported in the original studies to provide a fair and consistent comparison.

\subsection{Evaluation Metrics and Baseline Methods}

To evaluate the performance of our model, we employ two widely used metrics: Precision@$K$ (P@$K$) and Mean Reciprocal Rank@$K$ (MRR@$K$), with $K$ values set to 10 and 20. These metrics enable a robust assessment of both recommendation accuracy and ranking quality, which are essential for session-based recommendation systems. Specifically, P@$K$ measures the proportion of correctly recommended items within the top $K$ predictions. This metric focuses on the relevance of the recommended items within the immediate scope of user interaction, ensuring that the model prioritizes the most relevant items. As a complement, MRR@$K$ takes into account the rank of each correct recommendation, thereby assessing the ranking quality. By focusing on the position of relevant items within the top $K$ results, MRR@$K$ highlights the model's ability to present essential items early in the list, which enhances user satisfaction in real-world recommendation applications.

To provide a comprehensive evaluation, we compare our model with a variety of established baseline methods, covering RNN-based, GNN-based, and hypergraph-based models.

We select NARM~\citep{li17neural} as a representative RNN-based model as it incorporates attention mechanisms to capture user intent within sessions. NARM is recognized for its ability to handle sequential dependencies and has established a benchmark for sequence-based recommendation models by effectively modeling short-term interests. 

We also compare our model with three GNN-based models: SR-GNN~\citep{wu19session}, GCE-GNN~\citep{wang20global}, and COTREC~\citep{23}. SR-GNN utilizes session graphs to represent item interactions, allowing the model to capture complex relationships between items within sessions. GNN-based models have demonstrated substantial improvements in recommendation accuracy over traditional RNN models by leveraging graph structures. GCE-GNN, an advancement over SR-GNN, enhances session representation by integrating global context features, which further boosts the model's ability to generalize across diverse user sessions. The third GNN-based model, COTREC, introduces a self-supervised learning approach to capture both intra- and inter-session patterns using contrastive learning. This method has proven effective in reducing data sparsity issues, a common challenge in session-based recommendation systems.

In addition, we incorporate three hypergraph neural network (HGNN) models as baselines: DHCN~\citep{22}, SHARE~\citep{31}, and HIDE~\citep{10}. DHCN models user sessions as hypergraphs, connecting multiple items within a single hyperedge, which allows it to capture higher-order relationships that standard graph-based models often overlook. This capability is essential for applications where items are frequently co-purchased or viewed together, such as e-commerce platforms. SHARE employs a hypergraph attention mechanism, which selectively focuses on critical items within sessions, enhancing the model's precision in capturing session intent. Lastly, HIDE is designed to disentangle user intent within sessions by constructing a multi-type hypergraph structure. By distinguishing between different types of item interaction, HIDE can better represent the nuances of user preferences across various domains, making it a strong competitor in session-based recommendation tasks.

By comparing our model with these diverse baselines, we aim to highlight the advantages of our multi-channel approach, which combines knowledge graphs, hypergraphs, and line graphs to leverage external knowledge and session structure for improved recommendation quality. \rev{Although newer iterations of these models exist, these selected baselines serve as robust anchors to validate the specific architectural contributions of GraphFusionSBR, demonstrating the superiority of our unified fusion strategy.}.

\subsection{Recommendation Quality}

\begin{table*}[tb]
    \centering
    \caption{Recommendation quality comparison. The bold and underlined texts indicate the best and second-best results, respectively.}
    \resizebox{\textwidth}{!}{
    \begin{tabular}{c|c|c|ccc|cccc}
    \toprule
    \multirow{2}{*}{Dataset} & \multirow{2}{*}%
    {Metrics}
     & \multicolumn{1}{c|}{RNN} & \multicolumn{3}{c|}{GNN} & \multicolumn{4}{c}{HGNN} \\
    \cline{3-10}
     & & NARM & SR-GNN & GCE-GNN & COTREC & DHCN & SHARE & HIDE & GraphFusionSBR \\
    \midrule
    \multirow{4}{*}{Tmall} & P@10 & 26.34$\pm$0.30 & 23.71$\pm$0.17 & 27.58$\pm$0.09 & \underline{30.98$\pm$0.05} & 28.53$\pm$0.03 & 25.13$\pm$0.32 & 30.95$\pm$0.05 & \pmb{34.01$\pm$0.11} \\
     & MRR@10 & 15.23$\pm$0.20 & 13.31$\pm$0.09 & 14.85$\pm$0.09 & \underline{17.59$\pm$0.03} & 16.01$\pm$0.01 & 14.24$\pm$0.20 & 16.76$\pm$0.07 & \pmb{19.23$\pm$0.11} \\
     & P@20 & 31.09$\pm$0.28 & 28.25$\pm$0.17 & 32.74$\pm$0.11 & \underline{37.20$\pm$0.03} & 34.48$\pm$0.13 & 29.59$\pm$0.35 & 37.05$\pm$0.11 & \pmb{40.21$\pm$0.09} \\
     & MRR@20 & 15.51$\pm$0.18 & 13.62$\pm$0.08 & 15.21$\pm$0.09 & \underline{18.02$\pm$0.03} & 16.43$\pm$0.00 & 14.55$\pm$0.20 & 17.19$\pm$0.07 & \pmb{19.66$\pm$0.10} \\
     \midrule
     \multirow{4}{*}{RetailRocket} & P@10 & 64.96$\pm$0.12 & 64.28$\pm$0.22 & \underline{66.09$\pm$0.20} & 65.92$\pm$0.04 & 64.76$\pm$0.11 & 65.05$\pm$0.12 & 65.92$\pm$0.04 & \pmb{66.50$\pm$0.11} \\
     & MRR@10 & 47.74$\pm$0.05 & 47.26$\pm$0.17 & 48.09$\pm$0.09 & \underline{49.25$\pm$0.00} & 47.53$\pm$0.23 & 47.53$\pm$0.11 & 47.79$\pm$0.16 & \pmb{50.19$\pm$0.08} \\
     & P@20 & 70.20$\pm$0.32 & 69.57$\pm$0.14 & 71.48$\pm$0.15 & \underline{71.63$\pm$0.03} & 70.33$\pm$0.12 & 70.47$\pm$0.13 & 71.33$\pm$0.20 & \pmb{72.08$\pm$0.12} \\
     & MRR@20 & 48.11$\pm$0.04 & 47.63$\pm$0.17 & 48.47$\pm$0.08 & \underline{49.65$\pm$0.00} & 47.92$\pm$0.23 & 47.53$\pm$0.12 & 48.17$\pm$0.18 & \pmb{50.58$\pm$0.07} \\
     \midrule
     \multirow{4}{*}{KKBox} & P@10 &32.90$\pm$0.11 & 34.86$\pm$0.16 & 36.55$\pm$0.07 & 36.30$\pm$0.06 & 36.36$\pm$0.05 & 34.29$\pm$0.20 & \underline{37.02$\pm$0.03} & \pmb{37.15$\pm$0.08} \\
     & MRR@10 & 24.87$\pm$0.07 & \pmb{28.37$\pm$0.18} & 25.08$\pm$0.04 & 26.26$\pm$0.01 & 24.89$\pm$0.05 & 25.88$\pm$0.14 & 24.38$\pm$0.06 & \underline{26.49$\pm$0.15} \\
     & P@20 & 37.23$\pm$0.09 & 38.24$\pm$0.18 & 41.16$\pm$0.07 & 40.86$\pm$0.11 & 41.17$\pm$0.04 & 38.02$\pm$0.17 & \underline{41.91$\pm$0.05} & \pmb{42.09$\pm$0.07} \\
     & MRR@20 & 25.16$\pm$0.07 & \pmb{28.61$\pm$0.06} & 25.40$\pm$0.04 & 26.58$\pm$0.01 & 25.23$\pm$0.06 & 26.14$\pm$0.14 & 24.73$\pm$0.05 & \underline{26.83$\pm$0.15} \\
    \bottomrule
    \end{tabular}
    }
    \label{tab:performance}
\end{table*}

To assess the performance of our model's recommendation, we perform each experiment five times, recording the mean and standard deviation of all results, as shown in Table~\ref{tab:performance}. This approach helps to ensure the reliability and robustness of our model's performance across multiple runs.

Our analysis of the results reveals several vital observations. First, models leveraging graph structures generally outperform the RNN-based NARM. This consistent trend suggests that graph-based models, such as GNNs and HGNNs, are likely effective at capturing high-order relationships and collaborative filtering signals, which are crucial for understanding complex item interactions within sessions. NARM, while effective in sequence modeling, appears limited in its capacity to model these intricate relationships as effectively as graph-based methods.

Focusing on the three GNN-based recommendation models -- SR-GNN, GCE-GNN, and COTREC -- we observe that while COTREC achieves better results on e-commerce recommendation tasks, SR-GNN performs best for music recommendations. This performance difference could be attributed to the nature of the data and the specific modeling strengths of each approach. COTREC's advantage in e-commerce is likely due to its contrastive learning component, which enhances the model's ability to differentiate between relevant and irrelevant items, effectively capturing high-order item co-occurrence patterns typical in e-commerce sessions, where users often explore multiple similar or related items within a short period.
On the other hand, SR-GNN's strong performance in music recommendation scenarios could be due to its ability to capture sequential dependencies effectively, which are crucial in music streaming, where user preferences often follow specific genres, artists, or moods in sequence. Music sessions tend to have a more defined flow and coherence in terms of genre or artist similarity, making SR-GNN's simpler session graph architecture particularly effective. Additionally, SR-GNN's ability to represent item transitions directly in session graphs without contrastive learning may avoid the additional complexity introduced by self-supervised signals, which may not add as much value in the sequentially driven music context. Consequently, each model demonstrates strengths tailored to different session structures and recommendation patterns, making them more or less suitable for varying content types.

Among the hypergraph neural network (HGNN) models -- DHCN, SHARE, HIDE, and GraphFusionSBR -- our model, GraphFusionSBR, consistently performs the best. This performance boost can be attributed to GraphFusionSBR's integration of a knowledge graph, which enriches the item representations by incorporating additional contextual information. The knowledge graph channel enables GraphFusionSBR to leverage domain-specific relationships and semantic connections between items, thereby delivering more personalized and context-aware recommendations. This integration is particularly beneficial for capturing complex dependencies between items that are not necessarily sequential but are semantically related.

Furthermore, our model shows strong performance across all three datasets. These results demonstrate that the multi-channel architecture of GraphFusionSBR, which combines knowledge graphs, hypergraphs, and line graphs, effectively captures both high-order item interactions within sessions and external contextual knowledge. This architecture appears to be particularly advantageous in scenarios with diverse item categories and dynamic user interests, such as e-commerce and music streaming platforms.

In conclusion, GraphFusionSBR's superior performance in comparison to all baselines highlights the effectiveness of our multi-channel approach. Using a comprehensive knowledge graph and integrating it with session-based graph structures, our model excels at delivering accurate and contextually relevant recommendations, as demonstrated by its strong performance across various datasets and metrics.

\subsection{Ablation Study} \label{section:ablation-component}

\begin{table*}[tb]
    \centering
    \caption{Ablation study results on different datasets. The variants represent the model with specific components removed: NP (without Reverse Position Embedding), NKG (without Knowledge Graph Channel), NDKG (without KG Denoising Module), and NIEM (without Importance Extraction Module).}
    \resizebox{\textwidth}{!}{
    \begin{tabular}{lcccccccc}
        \toprule
        \multirow{2}{*}{Method} & \multicolumn{2}{c}{Tmall} & \multicolumn{2}{c}{RetailRocket} & \multicolumn{2}{c}{KKBox} \\
        \cline{2-7}
        & P@20(\%) & MRR@20(\%) & P@20(\%) & MRR@20(\%) & P@20(\%) & MRR@20(\%) \\
        \midrule
    GraphFusionSBR-NP & 40.19 & 19.63 & 69.77 & 48.64 & 38.18 & 17.58 \\
    GraphFusionSBR-NKG & 37.57 & 18.53 & 71.73 & 49.70 & 41.36 & 26.14 \\
    GraphFusionSBR-NDKG & 39.68 & 19.39 & 72.05 & 50.24 & 41.97 & 26.66 \\
    GraphFusionSBR-NIEM & 39.92 & 19.60 & 72.04 & 50.42 & 41.96 & 26.30 \\
    \midrule
    GraphFusionSBR & \pmb{40.21} & \pmb{19.66} & \pmb{72.08} & \pmb{50.58} & \pmb{42.09} & \pmb{26.83} \\
    \bottomrule
    \end{tabular}
    }
    \label{tab:ablation}
\end{table*}

To understand the contribution of each module within GraphFusionSBR, \rev{we compare GraphFusionSBR with the following variants (see Table~{\ref{tab:ablation}}):}
\begin{itemize}
    \item \rev{GraphFusionSBR-NP (No Position): removes the reverse position embedding module.}
    \item \rev{GraphFusionSBR-NKG (No Knowledge Graph): excludes the entire Knowledge Graph channel.}
    \item \rev{GraphFusionSBR-NDKG (No Denoised KG): uses the KG channel but removes the ``view generator'' denoising module, utilizing the full raw KG.}
    \item \rev{GraphFusionSBR-NIEM (No Importance Extraction): removes the Item Importance Extraction Module (IEM) from the Line Graph channel, treating all session interactions equally.}
\end{itemize}

We evaluated the performance of these variants in all three datasets, with the results reported in Table~\ref{tab:ablation}. We focus on Precision@20 (P@20) and Mean Reciprocal Rank@20 (MRR@20) as the primary metrics.

Our experiments reveal that each module contributes to the quality of recommendations, although the degree of impact varies depending on the dataset's characteristics. For the Tmall dataset, reverse position embedding has a relatively minor effect. This may be due to the nature of Tmall's recommendation patterns, which often suggest similar items within a category (e.g., viewing a necklace may lead to recommendations for other necklaces). Such browsing behavior suggests that sessions on Tmall do not involve significant shifts in user interest, which reduces the need for positional encoding. In contrast, removing the knowledge graph results in a substantial decrease in performance, indicating that the structured information provided by the knowledge graph plays a crucial role in enhancing the quality of recommendations. The absence of knowledge graph denoising and intra-session denoising also reduces accuracy, underscoring the importance of these denoising techniques in filtering irrelevant information and maintaining high-quality representations.

In the RetailRocket and KKBox datasets, reverse position embedding has a significant impact on performance, particularly for KKBox, where maintaining the item click order within sessions is crucial. This finding suggests that position information is more critical for accurately capturing user intent in domains where user preferences follow a more sequential or time-sensitive pattern (such as music playlists in KKBox). Additionally, knowledge graph denoising and intra-session denoising have a comparatively minor impact on these datasets. One possible reason for the limited effect of knowledge graph denoising in RetailRocket and KKBox is that these datasets contain relatively low noise after our knowledge graph preprocessing and filtering. Consequently, the denoising module has less scope to remove irrelevant information, resulting in a less pronounced impact. For intra-session denoising, the frequent and dynamic transitions in user interest observed in RetailRocket and KKBox may explain their reduced effectiveness in these cases, as user preferences are more varied and less susceptible to noise within a single session.

In general, the ablation study highlights that the contribution of each module depends on the characteristics of the dataset and user behavior patterns. The knowledge graph provides vital contextual information that improves recommendations for e-commerce applications, such as Tmall, where item categories remain relatively stable within sessions. Meanwhile, reverse position embedding proves essential for datasets with sequential dependencies, such as KKBox, to accurately represent session dynamics. The findings emphasize that incorporating global and session-specific information, combined with selective denoising, enables GraphFusionSBR to adapt effectively to various recommendation tasks and user behaviors.

\subsection{Hyperparameter Analysis} \label{hyperparameter-study}

\begin{table*}[tb]
    \centering
    \caption{The influence of the hyperparameter values. We set $l \in \{1,2,3,4,5\}$, $\lambda_1 \in \{0.001, 0.005, 0.01, 0.05, 0.1, 0.2, 0.5, 1\}$, and $K \in \{5, 10, 15, 20\}$. We show the minimum and maximum values within each pair of brackets.}
    \resizebox{\textwidth}{!}{
    \begin{tabular}{lcccccccc}
        \toprule
        \multirow{2}{*}{} & \multicolumn{2}{c}{Tmall} & \multicolumn{2}{c}{RetailRocket} & \multicolumn{2}{c}{KKBox} \\
        \cline{2-7}
        & P@20(\%) & MRR@20(\%) & P@20(\%) & MRR@20(\%) & P@20(\%) & MRR@20(\%) \\
        \midrule
    Varying $l$ & [39.15, 40.21] & [19.20, 19.66] & [71.89, 72.08] & [50.40, 50.58] & [41.16, 42.09] & [26.28, 26.83] \\
    Varying $\lambda_1$ & [38.24, 40.09] & [19.42, 20.09] & [69.02, 72.08] & [49.39, 50.58] & [39.14, 42.09] & [25.98, 26.83] \\
    Varying $K$ & [39.85, 40.21] & [19.29, 19.66] & [71.91, 72.08] & [50.35, 50.58] & [41.72, 42.09] & [26.23, 26.83] \\
    \bottomrule
    \end{tabular}
    }
    \label{tab:hyperpara}
\end{table*}

GraphFusionSBR comprises three critical hyperparameters: $l$, $\lambda_1$, and $K$, which regulate the number of model layers, the impact of contrastive learning, and the number of positive and negative samples utilized in training, respectively. We test the model's sensitivity to variations in these hyperparameters to evaluate its robustness and optimize its performance.

We set $l$ to $\{1,2,3,4,5\}$, $\lambda_1$ to $\{0.001, 0.005, 0.01, 0.05, 0.1, 0.2, 0.5, 1\}$, and $K$ to $\{5, 10, 15, 20\}$. Table~\ref{tab:hyperpara} summarizes the results, showing that GraphFusionSBR performs consistently well across a range of settings for these hyperparameters. This indicates a degree of robustness in the model's architecture, suggesting that it can maintain high effectiveness without requiring extensive hyperparameter tuning, which is a valuable characteristic for practical applications.

\begin{figure}[tb]
\centering
\begin{subfigure}[b]{0.32\textwidth}
    \centering
    \includegraphics[width=\textwidth]{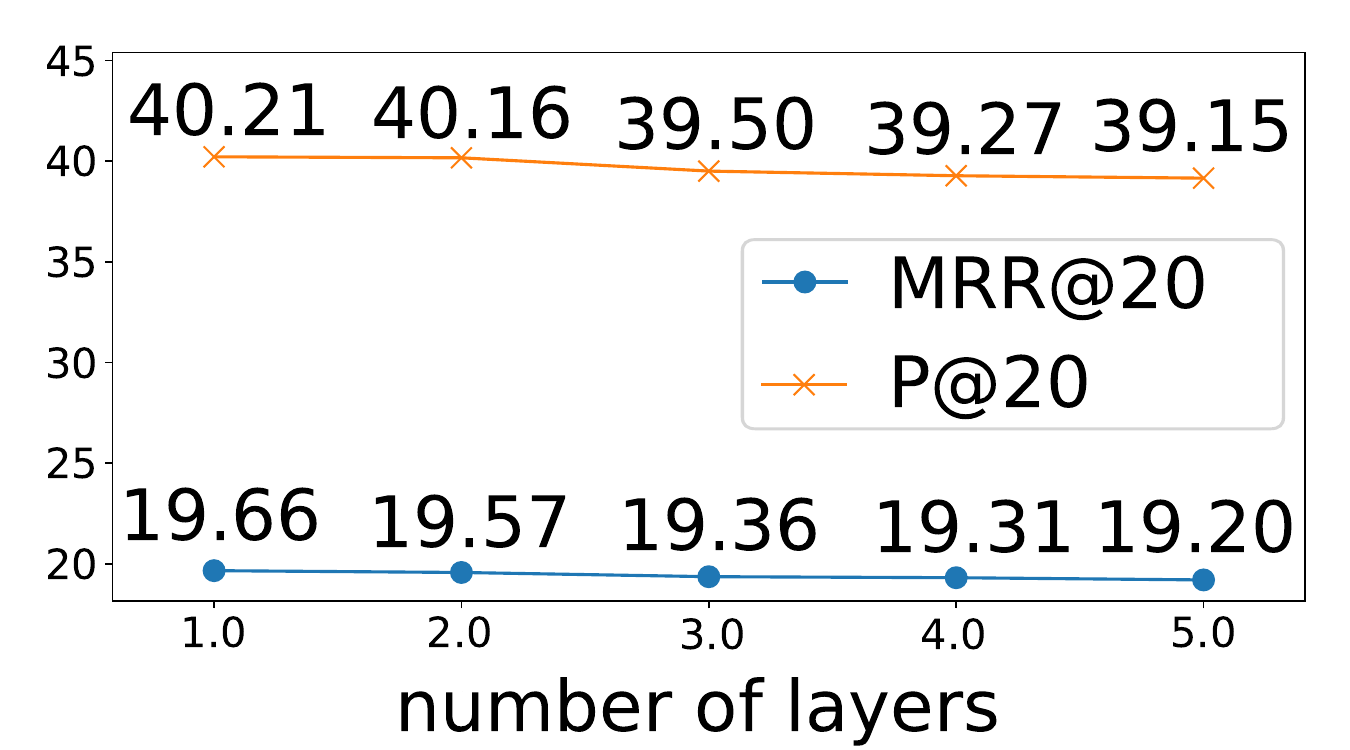}
    \caption{Tmall}
\end{subfigure}
\hfill
\begin{subfigure}[b]{0.32\textwidth}
    \centering
    \includegraphics[width=\textwidth]{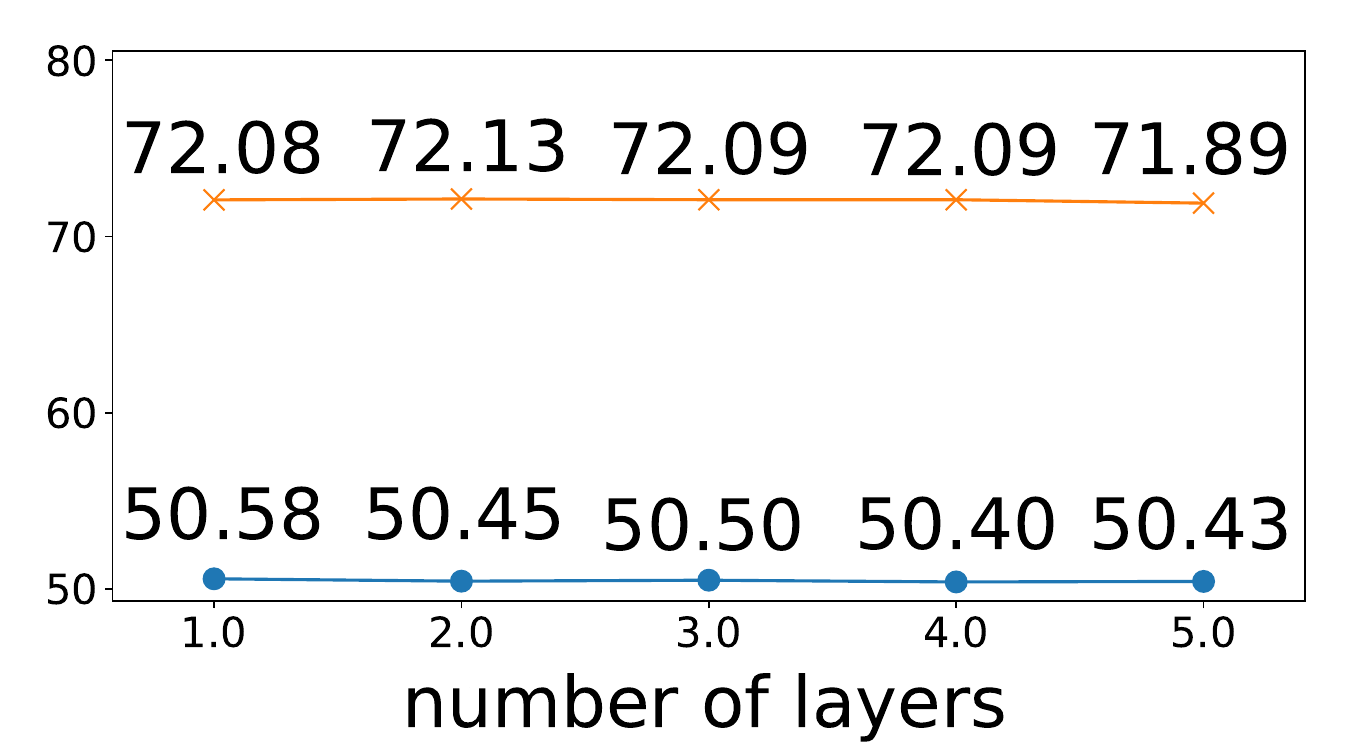}
    \caption{RetailRocket}
\end{subfigure}
\hfill
\begin{subfigure}[b]{0.32\textwidth}
    \centering
    \includegraphics[width=\textwidth]{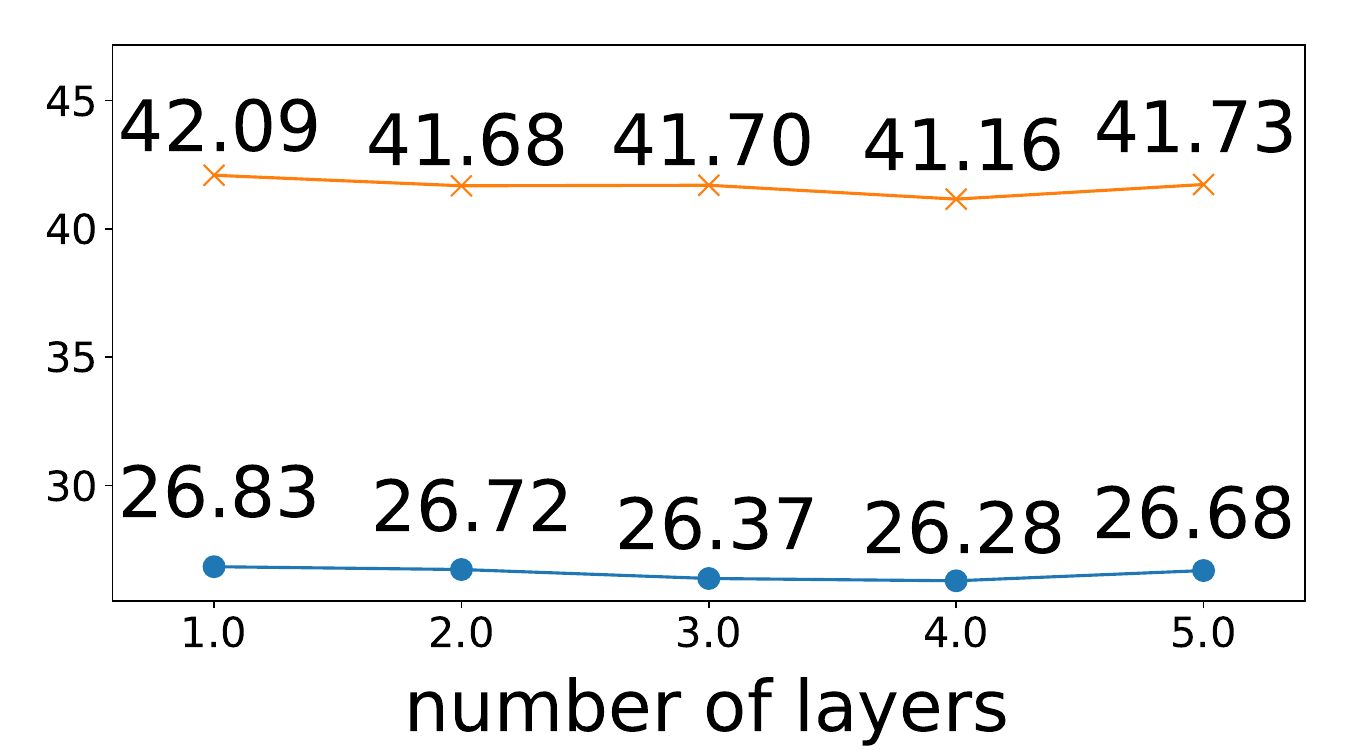}
    \caption{KKBox}
\end{subfigure}
\hfill
\caption{Effect of different layers on performance across datasets}
\label{fig:layer}
\end{figure}

Figure \ref{fig:layer} details the impact of the layer count $l$ on model performance in the three datasets, using Precision@20 (P@20) and Mean Reciprocal Rank@20 (MRR@20) as metrics.

In both the Tmall and KKBox datasets, the model achieves its best performance with a single layer, indicating that a shallow model structure is optimal in these cases. For Tmall, this may be due to the relatively straightforward nature of user purchasing behavior, where interactions are often direct and focused on specific categories. In such scenarios, adding depth does not significantly improve predictive power and may introduce unnecessary complexity. Similarly, in the KKBox dataset, users' music preferences tend to be highly subjective and sometimes less structured, suggesting that capturing session intent with deeper layers does not lead to performance gains. Instead, a single-layer model can effectively capture the relevant patterns.
In the RetailRocket dataset, performance remains relatively stable across different layer depths. This stability suggests that user browsing behavior on RetailRocket is less sensitive to model depth, likely due to the broader and more varied range of interactions within sessions. Given this consistency, we opted to use a single layer for RetailRocket in subsequent experiments to simplify model training without sacrificing performance.

\begin{figure}[tb]
\centering
\begin{subfigure}[b]{0.32\textwidth}
    \centering
    \includegraphics[width=\textwidth]{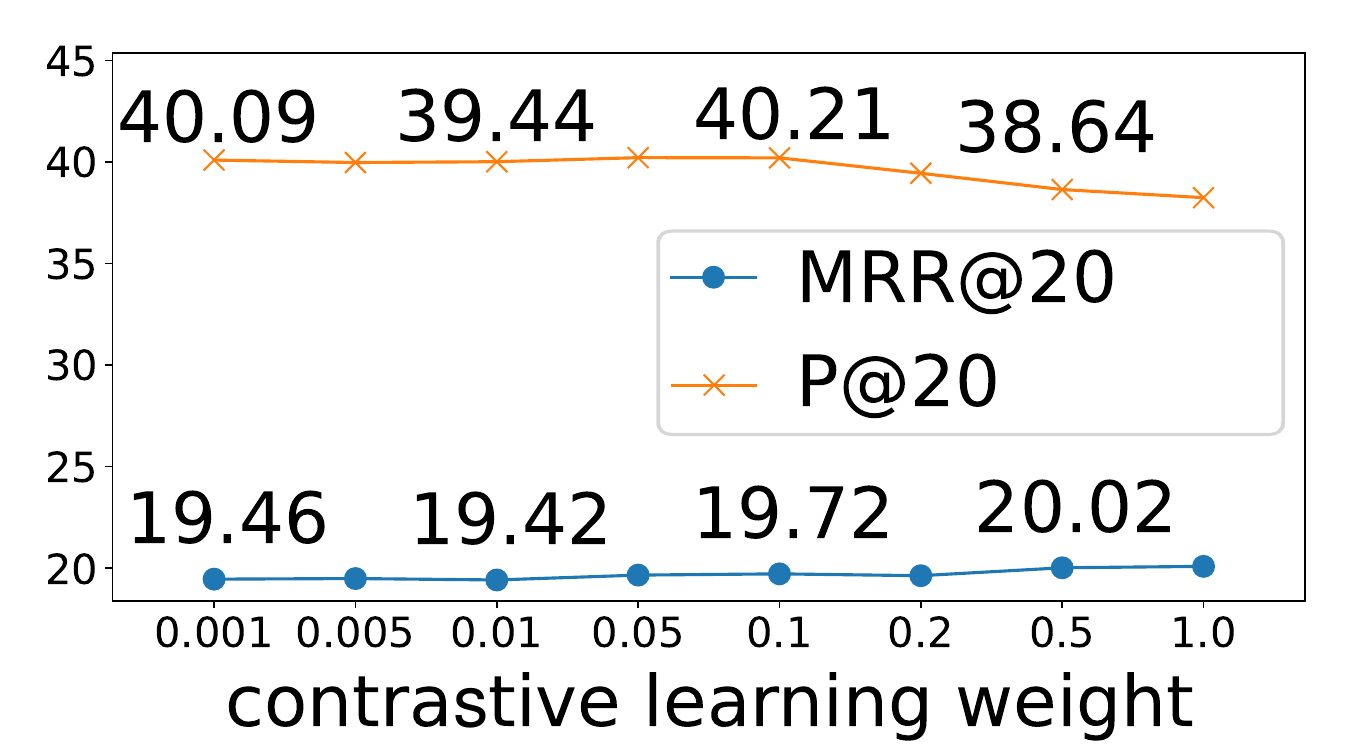}
    \caption{Tmall}
\end{subfigure}
\hfill
\begin{subfigure}[b]{0.32\textwidth}
    \centering
    \includegraphics[width=\textwidth]{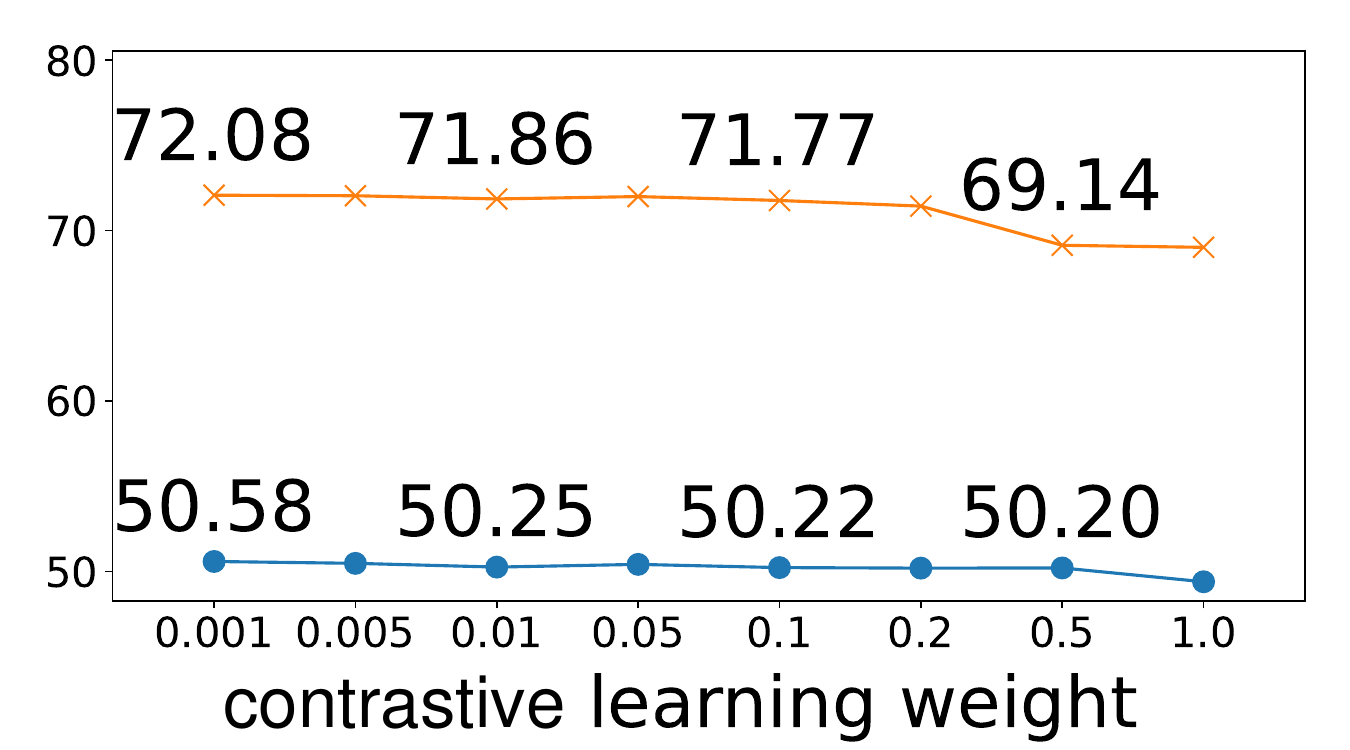}
    \caption{RetailRocket}
\end{subfigure}
\hfill
\begin{subfigure}[b]{0.32\textwidth}
    \centering
    \includegraphics[width=\textwidth]{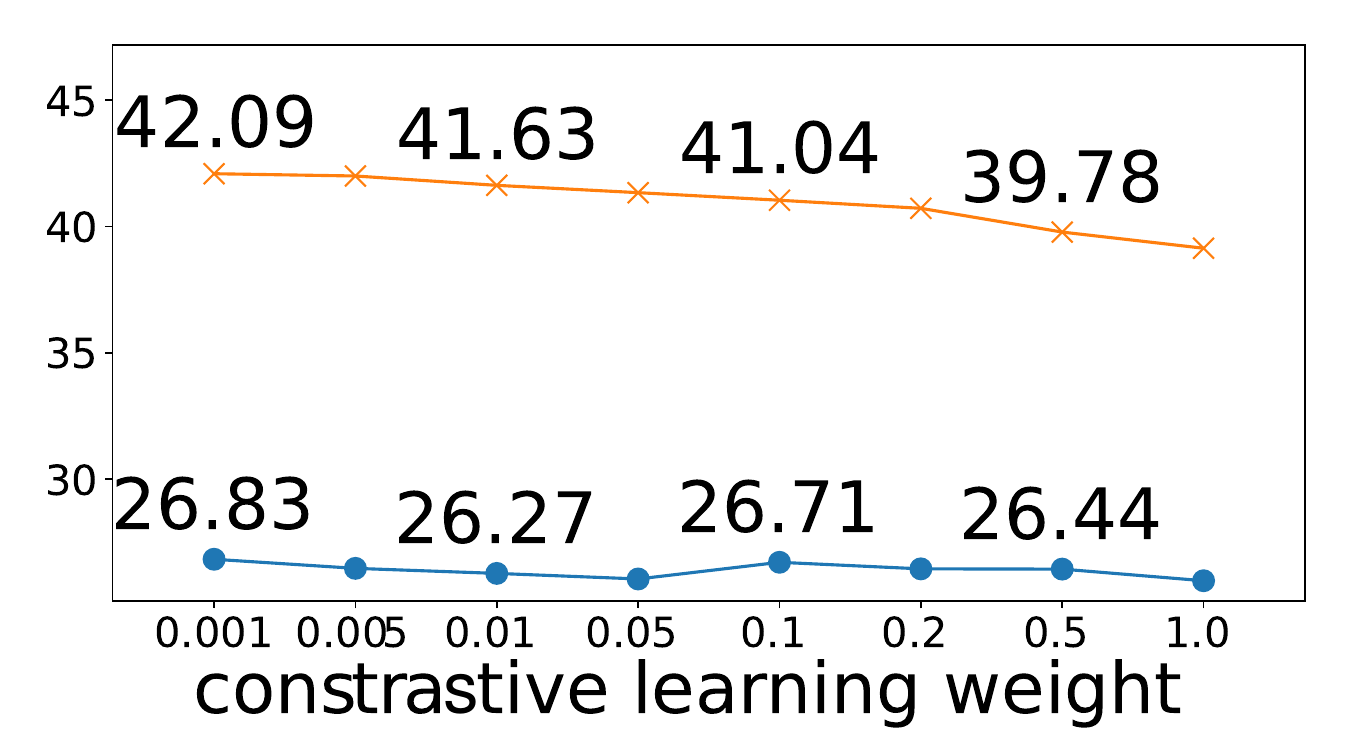}
    \caption{KKBox}
\end{subfigure}
\caption{Effect of different $\lambda_1$s on performance across datasets}
\label{fig:lambda}
\end{figure}

Regarding the effect of contrastive learning, the impact of $\lambda_1$ (the weight of contrastive learning) is shown in Figure~\ref{fig:lambda}. For Tmall, performance peaks at $\lambda_1 = 0.05$ (for P@20) and $\lambda_1=1$ (for MRR@20), while RetailRocket and KKBox show optimal results at a much lower value of $\lambda_1 = 0.001$. This suggests that contrastive learning has a more substantial effect on Tmall, possibly because Tmall sessions tend to involve more closely related items, where cross-session similarity information is valuable.

The importance of contrastive learning on Tmall, compared to RetailRocket and KKBox, can likely be attributed to distinct user interaction patterns and browsing behaviors specific to this e-commerce platform. Tmall users often exhibit high item similarity within sessions, especially in categories such as fashion and accessories, where contrastive learning effectively captures cross-session similarities and reinforces shared preferences, leading to more accurate recommendations. In contrast, RetailRocket's diverse platform services result in a broader end-user base with varied interests, reducing the impact of contrastive learning as sessions encompass a wider range of categories with less consistent patterns. Similarly, on KKBox, users' music preferences are highly individualized and influenced by personal tastes, moods, and cultural factors, introducing significant variability in session composition~\citep{bello2021cultural}. This diversity limits the effectiveness of contrastive learning in KKBox, as it becomes challenging to capture meaningful cross-session similarities between such unique listening behaviors.

\begin{figure}[tb]
\centering
\begin{subfigure}[b]{0.32\textwidth}
    \centering
    \includegraphics[width=\textwidth]{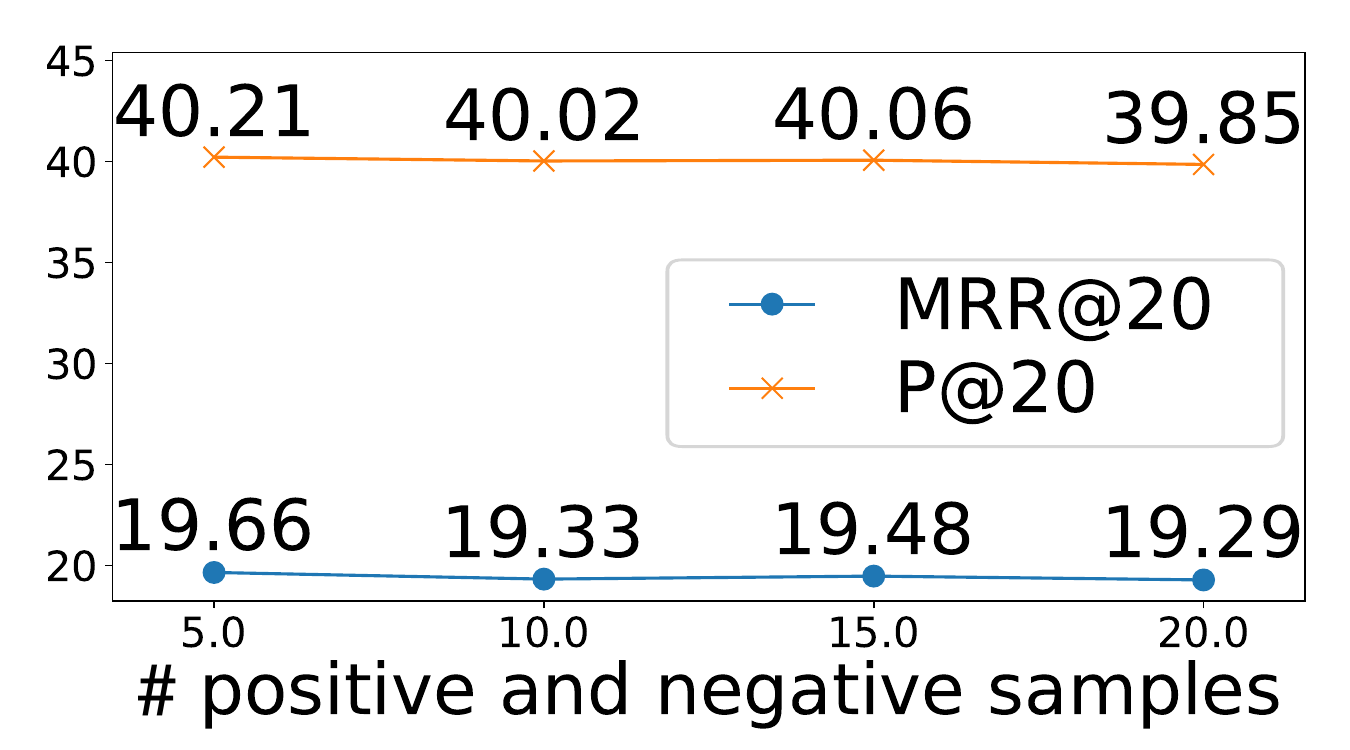}
    \caption{Tmall}
\end{subfigure}
\hfill
\begin{subfigure}[b]{0.32\textwidth}
    \centering
    \includegraphics[width=\textwidth]{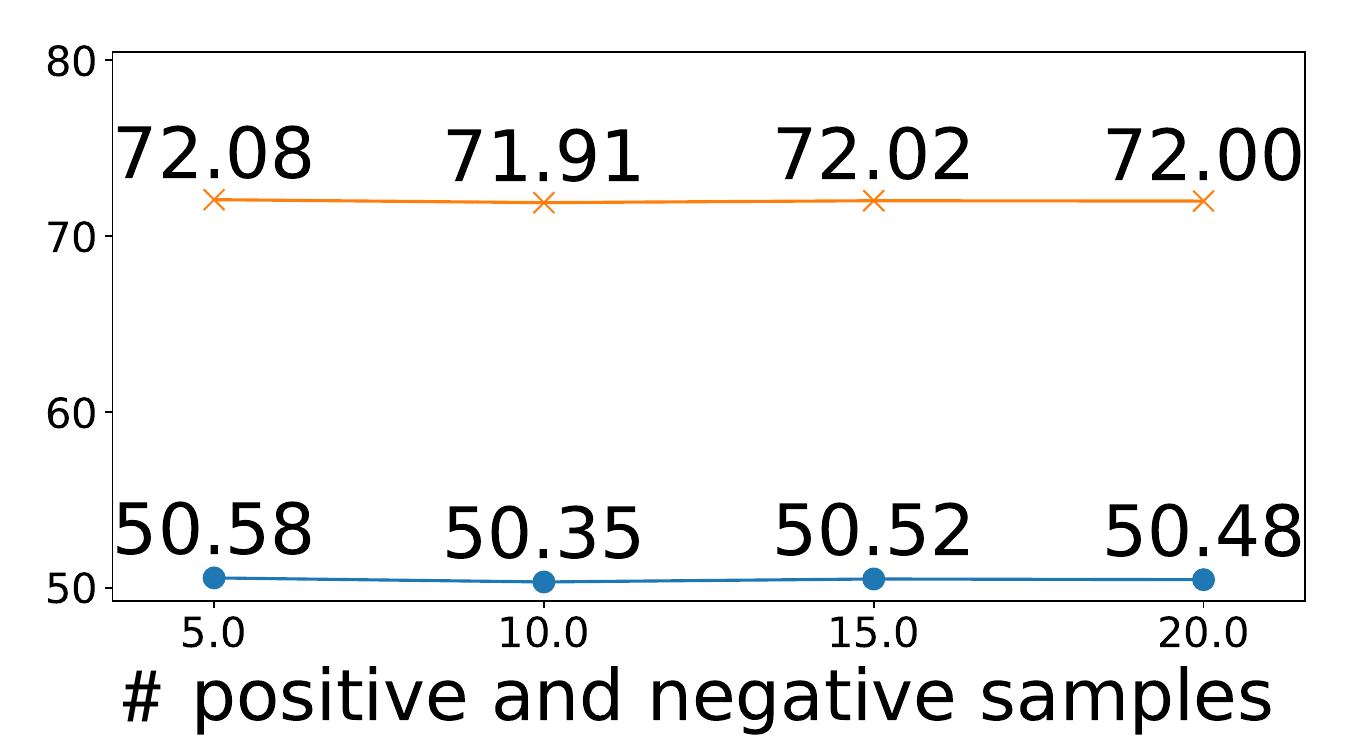}
    \caption{RetailRocket}
\end{subfigure}
\hfill
\begin{subfigure}[b]{0.32\textwidth}
    \centering
    \includegraphics[width=\textwidth]{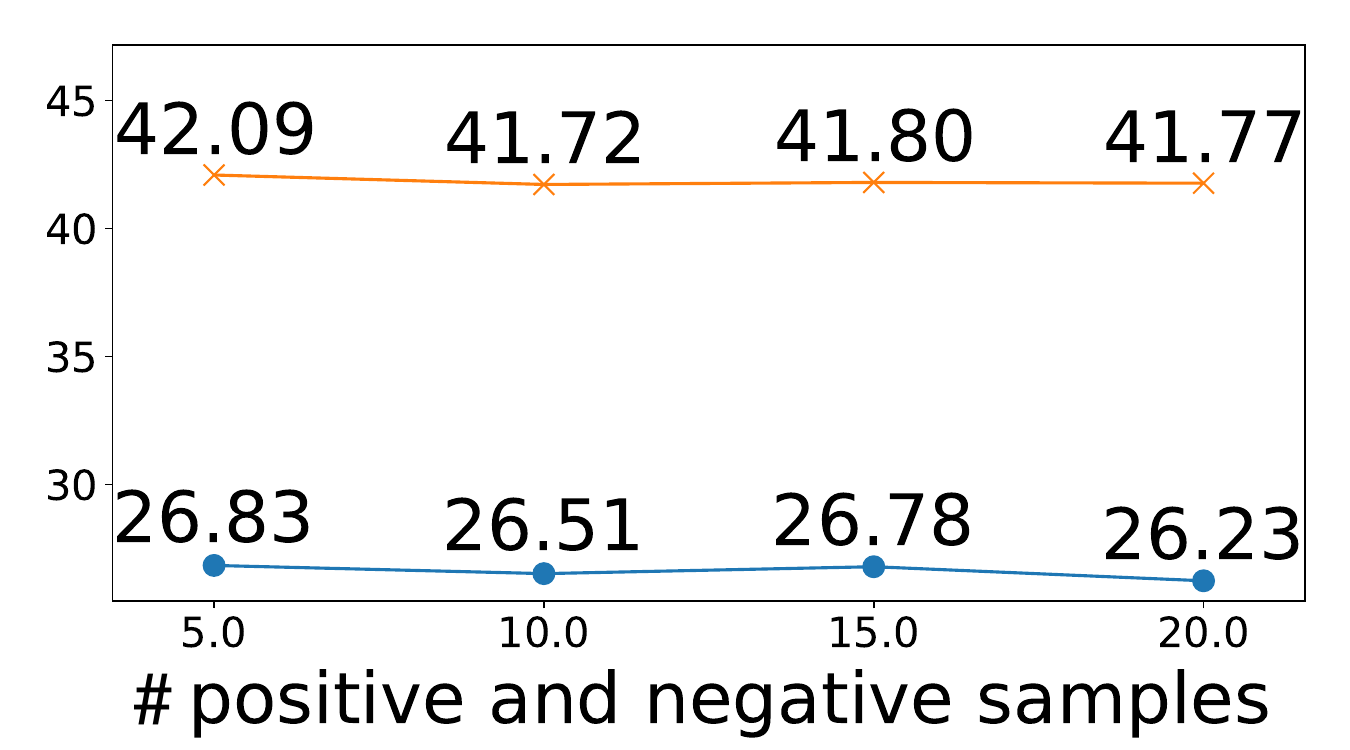}
    \caption{KKBox}
\end{subfigure}
\hfill
\caption{Effect of different $K$s on performance across datasets}
\label{fig:K}
\end{figure}

Figure \ref{fig:K} presents the effect of varying $K$ (the number of positive and negative samples) on the performance of the model in the three datasets. The model achieves its best performance across all datasets when $K$ is set to 5.

Using a smaller $K$ appears to yield a more concentrated and high-quality positive sample set, which may facilitate the model's ability to learn meaningful representations of user preferences. As $K$ increases, the size of both positive and negative sample sets increases, which can introduce noise and complicate the learning process. This can make it harder for the model to identify distinct user interests, particularly in datasets with limited interactions or smaller session sizes.

The consistent preference for $K = 5$ suggests that a smaller number of samples provides a balance between diversity and relevance in the sample sets. This allows the model to focus on highly relevant items without being overwhelmed by excessive or loosely related examples. Consequently, smaller $K$ values not only simplify training but also maintain robust performance across datasets, further highlighting the model's adaptability to different user behaviors and interaction types.

\section{Discussion} \label{sec:disc}

In this study, we present GraphFusionSBR, a session-based recommendation model that integrates three types of graph structures: hypergraphs, line graphs, and knowledge graphs. \rev{Crucially, unlike existing fusion methods that simply aggregate features, our multi-channel architecture incorporates explicit denoising mechanisms. Specifically, we employ a view generator to prune irrelevant semantic edges in the knowledge graph and an importance extraction module to filter noisy interactions within sessions. This design uses structured session data and contextual knowledge to effectively capture both intra-session and inter-session item relationships. By combining these denoised graph types, GraphFusionSBR creates a more comprehensive and robust representation of user behavior.} The experimental results on three datasets (Tmall, RetailRocket, and KKBox) demonstrate that this denoising-driven design leads to more accurate recommendations compared to the state-of-the-art SBR methods across most datasets. Notably, GraphFusionSBR's use of refined knowledge graphs provides additional contextual layers that boost recommendation relevance, especially on datasets where user interactions are more structured and category-focused, such as Tmall.

Despite the promising results of our proposed recommendation model, there is still room for improvement in specific scenarios. First, the knowledge information used in this study was limited to textual attributes, such as item categories and user demographics. Integrating other types of data, such as images, could allow the model to build a richer multimodal knowledge base. This would be especially useful in domains like fashion, where visual features often significantly influence user preferences. Incorporating such modalities could allow GraphFusionSBR to generate recommendations that more accurately reflect the aesthetic and sensory preferences of users, thus increasing the relevance and appeal of the recommendations.
Second, incorporating additional contextual information, such as date, weather, and location, could further enhance the model's ability to make relevant and timely recommendations. For example, a user's shopping behavior may vary significantly depending on seasonal factors or even weather conditions, which are particularly influential in domains like fashion and food. By incorporating these contextual dimensions, GraphFusionSBR can dynamically adjust recommendations in response to changing external factors, resulting in a more personalized experience for users. Context-aware recommendations would make the model more versatile and applicable in a wider range of scenarios, particularly in domains where situational factors heavily influence user preferences. The extra information can be seamlessly integrated into GraphFusionSBR by adding additional channels for contrastive learning.

Overall, while GraphFusionSBR demonstrates a robust framework for session-based recommendations, exploring these additional avenues could enable further improvements in recommendation quality, relevance, and personalization. Future work could focus on integrating multimodal and context-sensitive data sources, as well as enhancing temporal modeling, to create even more accurate and adaptable recommendation systems.

\section*{Acknowledgment and GenAI Usage Disclosure}

We acknowledge support from National Science and Technology Council of Taiwan under grant number 113-2221-E-008-100-MY3. 
The authors used ChatGPT and Gemini to improve language and readability. The authors reviewed and edited the content as needed and take full responsibility for the content of the publication.

\bibliographystyle{unsrt}  
\bibliography{ref}  

\begin{appendices}
\appendix

\section{Standard Hypergraph Convolution} \label{app:hyper-conv}

In the hypergraph channel, the degree of each node $v_i$ and hyperedge $e$ is defined as:

\begin{equation} \label{eq:hg-degree}
D_{ii}=\sum_{\epsilon=1}^{M}W_{\epsilon\epsilon}H_{i\epsilon}, \quad B_{ee}=\sum_{i=1}^{N}H_{i\epsilon}
\end{equation}

Based on these degree matrices, the hypergraph convolution at layer $l$ propagates information as follows:

\begin{equation} \label{eq:hg-conv}
X_{h}^{(l+1)} = D^{-1}HWB^{-1}H^{T}X_{h}^{(l)},
\end{equation}
where $H$ is the incidence matrix and $W$ is the hyperedge weight matrix.

\section{Graph Attention Network (GAT) Formulation} \label{app:gat}

The knowledge graph encoder utilizes GAT to update node representations. For a node $i$ and its neighbor $j$, the attention coefficient $\alpha_{ij}$ is computed as:

\begin{equation} \label{eq:gat-att-coef}
\alpha_{ij} = \frac{\exp(\text{LeakyReLU}(a^T [Wx_i || Wx_j]))}{\sum_{k \in N_i} \exp(\text{LeakyReLU}(a^T [Wx_i || Wx_k]))}
\end{equation}

The final representation is updated by aggregating neighbors:

\begin{equation} \label{eq:gat-node-embedding}
x_i^{(l+1)} = \sigma \left( \sum_{j \in N_i} \alpha_{ij} W x_j^{(l)} \right)
\end{equation}

\section{TransR Scoring Function} \label{app:transr-score}

To model the semantic relationships in the Knowledge Graph, we adopt the TransR scoring function. The plausibility score of a triple $(h,r,t)$ is defined as:

\begin{equation} \label{eq:triple-score}
g(h,r,t) = || W_r e_h + e_r - W_r e_t ||_2^2,
\end{equation}
where $W_r$ projects entities into the relation space.

\section{Training Algorithm} \label{app:train-algo}

\begin{algorithm}[H]
\LinesNumbered
    \KwIn{$X^{(0)}_h:$ hypergraph embedding,\ $X^{(0)}_k:$ knowledge graph embedding,\ $X^{(0)}_l:$ line graph embedding,\ $\mathcal{G}_k:$ knowledge graph,\ $\mathcal{G}_h:$ hypergraph,\ $\mathcal{G}_l:$\ line graph, $Y:$ true label set}

    \For{\textnormal{every epoch}}{
        \tcc{Knowledge Graph Channel}
        Find all triples ($h$, $r$, $t$) in $\mathcal{G}_k$\;
        Randomly select a fake tail $t^-$ for each triple\;
        Compute $g(h, r, t^-)$ and $g(h, r, t^+)$\;
        $L_{KG}$ = $-\log \sigma(g(h, r, t^-) - g(h, r, t^+))$\;
        Using $L_{KG}$ to update model's weights and embeddings\;

        $\hat{\mathcal{G}_k}$ = KGDenoising($\mathcal{G}_k$)\;
        $X_k, \Theta_k$ = KGEncoder($\hat{\mathcal{G}_k}$, $X^{(0)}_k$)\;

        \tcc{Hypergraph Channel}
        $X_h, \Theta_h$ = HGEncoder($\mathcal{G}_h$, $X^{(0)}_h$)\;

        \tcc{Recommendation Task}
        $\hat{Z}$ = $(\Theta_h || \Theta_k)^T (X_{h} || X_{k})$\;
        $\hat{Y}$ = $\text{Softmax}(\hat{Z})$\;
        $L_{r}$ = CrossEntropy($\hat{Y}$, $Y$)\;

        \tcc{Line Graph Channel \& Co-training}
        $\Theta^{(0)}_l$ = IEM($X^{(0)}_h$)\;
        $\Theta_l$ = LGEncoder($\mathcal{G}_l$, $\Theta^0_l$)\;
        $c^{+}_l, c^{-}_l, c^{+}_h, c^{-}_h$ = Co-training($\Theta_l$, $\Theta_h$, $X^{(0)}_h$, $X_h$)\;
        $L_{ssl}$ = InfoNCE($\Theta_l$, $X^{\text{last}}_l$, $c^{+}_l$, $c^{-}_l$) + InfoNCE($\Theta_h$, $X^{\text{last}}_h$, $c^{+}_h$, $c^{-}_h$)\;
        Using $L_{r}$ and $L_{ssl}$ to update model's weights and embeddings\;
    }
\caption{GraphFusionSBR training algorithm} \label{alg:train}
\end{algorithm}

\end{appendices}
\end{document}